\documentclass[reprint,superscriptaddress,amsmath,amssymb,aps]{revtex4-2}
\usepackage{CJK}
\usepackage[utf8]{inputenc}
\usepackage{graphicx}
\usepackage{dcolumn}
\usepackage{bm}
\usepackage{tikz}
\usepackage{todonotes}
\usepackage{bm}
\usepackage{hyperref}

\usetikzlibrary{patterns}
\usepackage{mathrsfs}

\definecolor{MyGreen}{RGB}{54,165,54}

\newcommand{\comment}[1]{}
\newcommand{\NN}[1]{next\textbf{nano}#1}
\begin{document}

\title{Impact of random alloy fluctuations on the carrier distribution in multi-color (In,Ga)N/GaN quantum well systems}
\author{Michael O'Donovan}
\affiliation{Tyndall National Institute, University College Cork, Cork, T12 R5CP, Ireland}
\affiliation{Department of Physics, University College Cork, Cork, T12 YN60, Ireland}

\author{Patricio Farrell}
\affiliation{Weierstrass Institute (WIAS), Mohrenstr. 39, 10117 Berlin, Germany}

\author{Julien Moatti}
\affiliation{Inria, Univ. Lille, CNRS, UMR 8524 - Laboratoire Paul Painlev{\'e}, F-59000 Lille, France}

\author{Timo Streckenbach}
\affiliation{Weierstrass Institute (WIAS), Mohrenstr. 39, 10117 Berlin, Germany}

\author{Thomas Koprucki}
\affiliation{Weierstrass Institute (WIAS), Mohrenstr. 39, 10117 Berlin, Germany}

\author{Stefan Schulz}
\affiliation{Tyndall National Institute, University College Cork, Cork, T12 R5CP, Ireland}
\affiliation{Department of Physics, University College Cork, Cork, T12 YN60, Ireland}


% ...
\begin{abstract}
    In this work, we study the impact that random alloy fluctuations have on the distribution of electrons and holes across the active region of a (In,Ga)N/GaN multi-quantum well based light emitting diode (LED). To do so, an atomistic tight-binding model is employed to account for alloy fluctuations on a microscopic level and the resulting tight-binding energy landscape forms input to a quantum corrected drift-diffusion model. Here, quantum corrections are introduced via localization landscape theory and we show that when \emph{neglecting alloy disorder} our established theoretical framework yields results very similar to commercial software packages that employ a self-consistent Schr\"odinger-Poisson-drift-diffusion solver; this provides validation of the developed quantum corrected transport model. Similar to experimental studies in the literature, we have focused on a multi-quantum system where two of the three wells have the same In content while the third well differs in In content. By changing the order of wells in this `multi-color' quantum well structure and looking at the relative radiative recombination rates of the different emitted wavelengths, we (i) gain insight into the distribution of carriers in such a system and (ii) can compare our findings to trends observed in experiment. 
    Our results indicate that the distribution of carriers depends significantly on the treatment of the quantum well microstructure. For instance, when including random alloy fluctuations and quantum corrections in the simulations, the calculated trends in the relative radiative recombination rates as a function of the well ordering are consistent with previous experimental studies. 
    However, the results from the widely employed virtual crystal approximation contradict the experimental data. Our calculations clearly demonstrate that when accounting for random alloy fluctuations in the simulations,  no further ad-hoc modifications to the transport model are required, in contrast to previous studies neglecting alloy disorder. Overall, our work highlights the importance of a careful and detailed theoretical description of the carrier transport in an (In,Ga)N/GaN multi-quantum well system to ultimately guide the design of the active region of III-N-based LED structures. 
\end{abstract}
\maketitle


\section{Introduction}



At the heart of modern light emitting diodes (LEDs) operating in the blue to violet spectral region are (In,Ga)N/GaN multi-quantum well (MQW) systems~\cite{Hump2008}. While the efficiency of these LEDs is and can be very high, further efficiency gains will still directly reduce the cost of operating such LEDs. Moreover, extending efficient operation of (In,Ga)N-based LEDs into the green to red spectral range is a topic of current research interest~\cite{QuLiACSP2019,ScBiPhsStatSol2013, deMaPecPRL2016,DuBaJAP2020,HwHaAPE2014}. To achieve all this, understanding the fundamental electronic, optical and transport properties of (In,Ga)N-based MQW systems is of central importance to guide design of (In,Ga)N-based LEDs with new and improved capabilities. While experimental and theoretical studies on the electronic~\cite{CaSc2011,ChODo2021,TaCaRSC2016} and optical~\cite{JoTeAPL2017,DaScJAP2016} properties of such systems have already revealed that these properties are significantly impacted by alloy fluctuation induced carrier localization effects, the impact of alloy disorder on the carrier transport has only been targeted recently~\cite{ODoLu2021,MiOD2021_JAP,OdFaOQE2022,LyLhPhysRevApplied2022,BrMa2015,LiPi2017}. Here, studies are ranging from fully atomistic quantum mechanical approaches~\cite{ODoLu2021} up to modified continuum-based models~\cite{MiOD2021_JAP,OdFaOQE2022,LyLhPhysRevApplied2022,BrMa2015,LiPi2017}. 

Recently, we have developed a three-dimensional (3-D) multiscale simulation framework that connects atomistic tight-binding theory with a modified, quantum corrected drift-diffusion (DD) solver. The framework has been employed to investigate uni-polar carrier transport and it was found that alloy fluctuations result in an increase in carrier transport of electrons (in an $n$-$i$-$n$ system)~\cite{MiOD2021_JAP}, but decreases transport in the case of holes (in a $p$-$i$-$p$ systems)~\cite{OdFaOQE2022}. In the present work we extend this scheme to investigate the active region of (In,Ga)N-based MQW LED structures (thus $p$-$i$-$n$ systems) and study how carriers distribute across the active region. In general, understanding the carrier distribution can help to guide maximizing the efficiency in an LED, since ideally the carriers shall be distributed evenly across the entire MQW region so that all QWs will contribute to emission~\cite{ZhZhJAP2011}.
However, previous experimental studies on carrier distribution in (In,Ga)N/GaN MQW systems have indicated that mainly the well closest to the $p$-doped contact side  contributes to the light emission process~\cite{BGaPhysStatSol_2011,AuGrAPL2008,LiRyAPL2008,ZhZhJAP2011}. These samples were specifically designed to gain insight into the carrier distribution inside the active region of an LED.

Overall, this has been attributed to a sequential filling of the QWs, resulting in a high hole density only in the $p$-side QW. To establish accurate carrier transport models the trends found in the experimental studies of Refs.~\cite{BGaPhysStatSol_2011,AuGrAPL2008,LiRyAPL2008,ZhZhJAP2011} need to be captured. Previous theoretical studies have reproduced the experimentally observed behaviour, however this required (i) treating bound carriers in a quantum mechanical picture, (ii) softening of the QW barrier interface to account for tunneling effects, (iii) distinguishing between continuum and bound carriers in the carrier transport model (multi-population model), and (iv) allowing for scattering between the different populations~\cite{RoWiSPIE2017}. But, the impact of alloy disorder is basically neglected in this advanced but also complex carrier transport model.



In this paper we show that when employing our quantum corrected 3-D simulation framework that accounts for random alloy fluctuations, the experimentally observed trends are captured, without introducing for instance a multi-population scheme. This highlights that our developed solver presents an ideal starting point for future device design studies. 

To highlight clearly the impact that random alloy fluctuations have on the carrier distribution in the active region of an (In,Ga)N-based LED, we use as a reference point a virtual crystal approximation (VCA) which effectively can be described by a 1-D model. The benefit of this is twofold. Firstly, this enables us to compare directly the outcomes of our quantum corrected model with results from 1-D commercial software simulations; 
commercial software packages often employ a standard Schr\"odinger-Poisson-DD solver, which is numerically very costly and therefore unfeasible in large 3-D transport simulations. This motivates the need for an alternative implementation of quantum corrections.
Secondly, and building on this benchmark, alloy fluctuations can be included in the calculations, revealing clearly their impact on the results.
Our studies show, and when using the same input parameter set, only the model accounting for random alloy fluctuations produces trends that are consistent with the experimental data. The widely employed VCA yields results that are in contradiction with the experimental data, thus indicating that radiative recombination stems mainly from the well \emph{furthest away} from the $p$-side. Overall, this highlights (i) that alloy fluctuations are essential to achieve an accurate description of the carrier transport and (ii) have to be taken into account when theoretically guiding the design of energy efficient III-N light emitters. 




The paper is organized as follows: in Section~\ref{sec:TheoryStructs} we outline the model structure used for calculations and briefly summarise some of the literature experimental data from Ref.~\cite{BGaPhysStatSol_2011}. The theoretical framework which we use is summarized in Section~\ref{sec:Theory}. Our results are discussed in Section~\ref{sec:Results}. Finally Section~\ref{sec:conclusion} presents our conclusions.

\section{Model MQW structures and literature experimental findings}\label{sec:TheoryStructs}

\begin{figure}
    \centering
    \includegraphics[width=0.95\columnwidth]{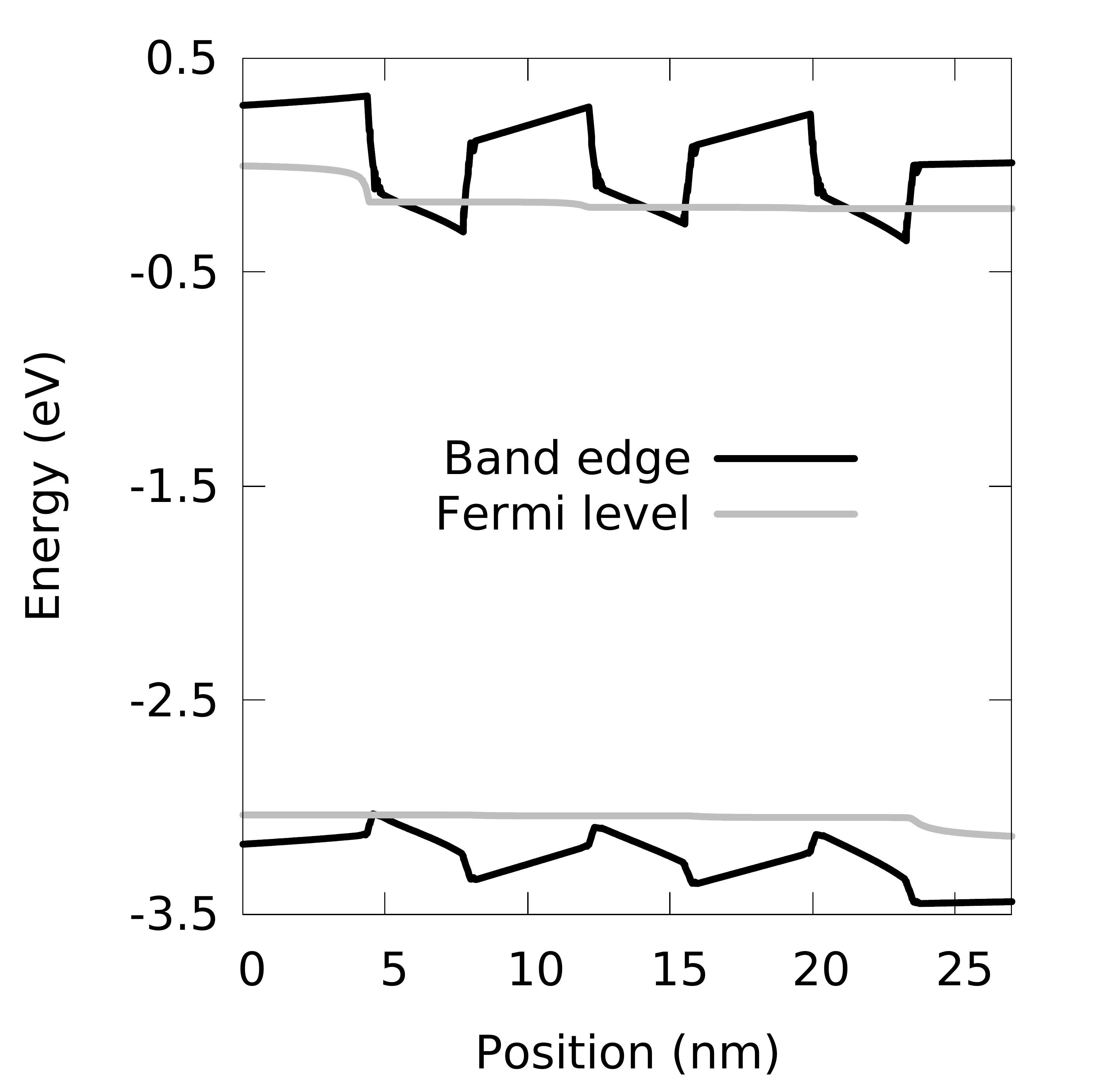}
    \caption{Conduction and valence band edges (black) along with the quasi-Fermi energies for electrons and holes (grey) in an (In,Ga)N/GaN multi-quantum well system described in virtual crystal approximation. The band edge profile and the quasi Fermi levels are shown at a current density of $50\text{ A/cm}^2$. The leftmost (In,Ga)N quantum well  contains 12.5\% indium while the other two (In,Ga)N wells (centre and right) contain 10\% indium.}
    \label{fig:BEs_1DVCA}
\end{figure}

To investigate the carrier distribution in (In,Ga)N/GaN MQW systems we proceed similar to experimental studies in the literature~\cite{BGaPhysStatSol_2011,LiRyAPL2008} and target MQW systems where one of the wells in the MQW stack has a slightly higher In content compared to the remaining wells. 
In our case, we study MQW systems with three (In,Ga)N/GaN wells. Here two are In$_{0.1}$Ga$_{0.9}$N (``shallow'') wells and one is an In$_{0.125}$Ga$_{0.875}$N (``deep'') QW. These QWs are 3 nm wide and separated by 5 nm GaN barriers. The band edge profile of such a system along the transport ($c$-) direction, using a VCA, is shown in Fig.~\ref{fig:BEs_1DVCA} at a current density of 50 A/cm$^2$.

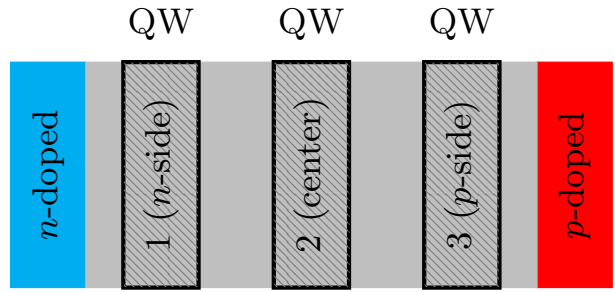
\begin{figure}[t]
    \centering
    \begin{tikzpicture}
        \filldraw[cyan, thin] (0,0) rectangle (1,3);
        \node[scale=1.5,rotate=90] at (0.5,1.5) {$n$-doped};
        \filldraw[red, thin] (7,0) rectangle (8,3);
        \filldraw[lightgray, thin] (1,0) rectangle (7,3);
        \node[scale=1.5,rotate=90] at (7.5,1.5) {$p$-doped};
        \draw[black, ultra thick]  (1.5,0) rectangle (2.5,3);
        \draw[pattern=north west lines, pattern color=gray] (1.5,0) rectangle (2.5,3);
        \node[scale=1.5,rotate=90] at (2,1.5) {1 ($n$-side)};
        \node[scale=1.5] at (2,3.5) {QW};
        \draw[black, ultra thick]  (3.5,0) rectangle (4.5,3);
        \draw[pattern=north west lines, pattern color=gray] (3.5,0) rectangle (4.5,3);
        \node[scale=1.5,rotate=90] at (4,1.5) {2 (center)};
        \node[scale=1.5] at (4,3.5) {QW};
        \draw[black, ultra thick]  (5.5,0) rectangle (6.5,3);
        \draw[pattern=north west lines, pattern color=gray] (5.5,0) rectangle (6.5,3);
        \node[scale=1.5,rotate=90] at (6,1.5) {3 ($p$-side)};
        \node[scale=1.5] at (6,3.5) {QW};
    \end{tikzpicture}
    \caption{Schematic illustration of multi-quantum well system. The $n$-doped region is shown in cyan, the $p$-doped is in red and undoped regions are in grey. The quantum wells are numbered starting from the $n$-side.}
    \label{fig:schematic_of_QWs}
\end{figure}



In the following, we investigate the carrier transport properties in two settings: (i) on an atomistic level accounting for random alloy fluctuations and (ii) in the frame of a VCA thus neglecting alloy fluctuations. In the latter VCA case (ii), at a given $z$-position (along the \emph{c}-direction), there is no variation in material properties within the growth plane ($c$-plane). This assumption is also made in the widely used 1-D transport simulations on (In,Ga)N MQWs.
\begin{figure*}[t]
    \centering
    \begin{tabular}{|c|c|c|}
    \hline
     \large \textbf{(a)} Valence band edge \normalsize & \large \textbf{(b)} Conduction band edge \normalsize & \large \textbf{(c)} Radiative recombination  \normalsize \\
    \hline
        \includegraphics[width=0.3\textwidth]{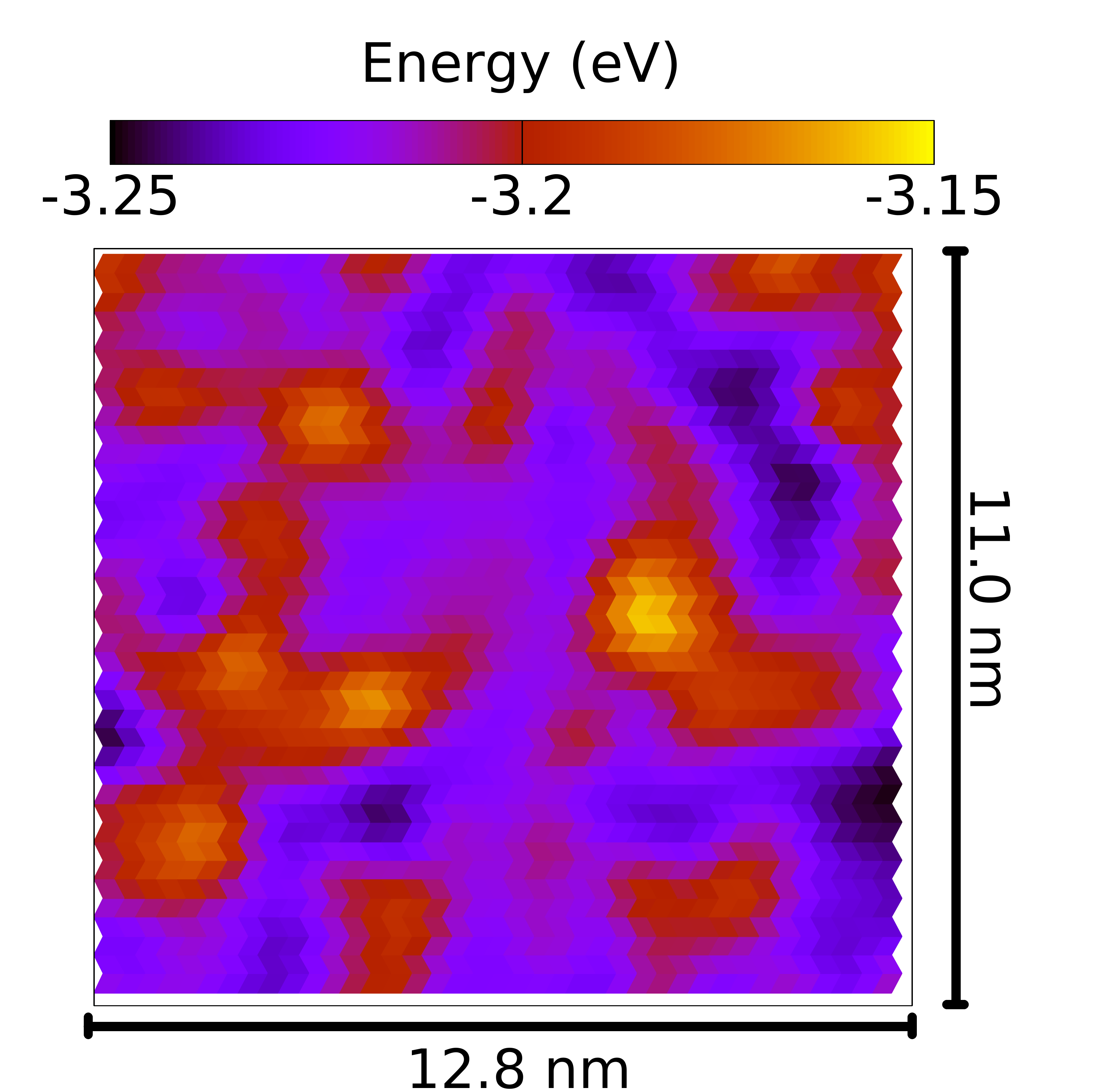} & 
        \includegraphics[width=0.3\textwidth]{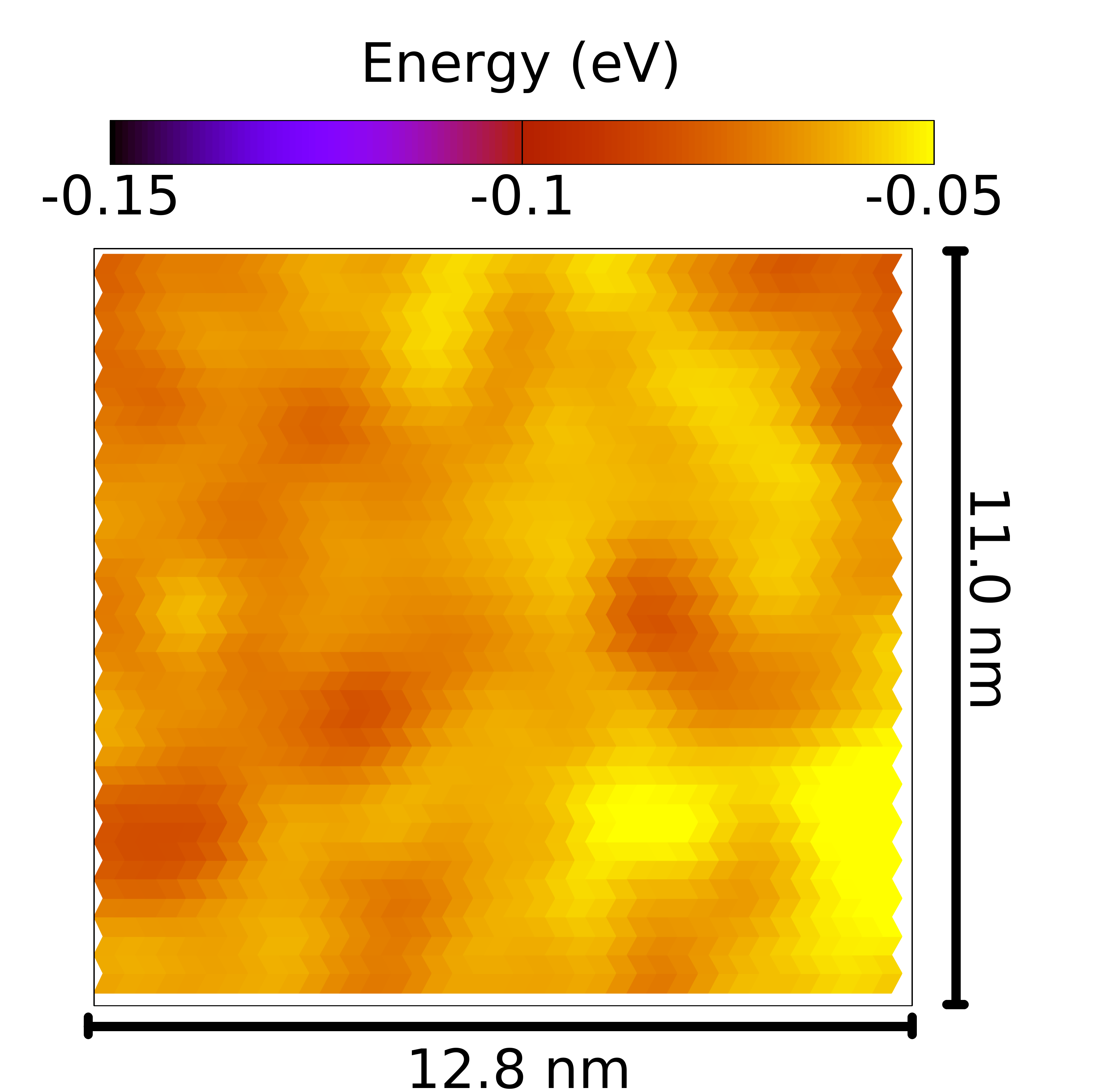} &
        \includegraphics[width=0.3\textwidth]{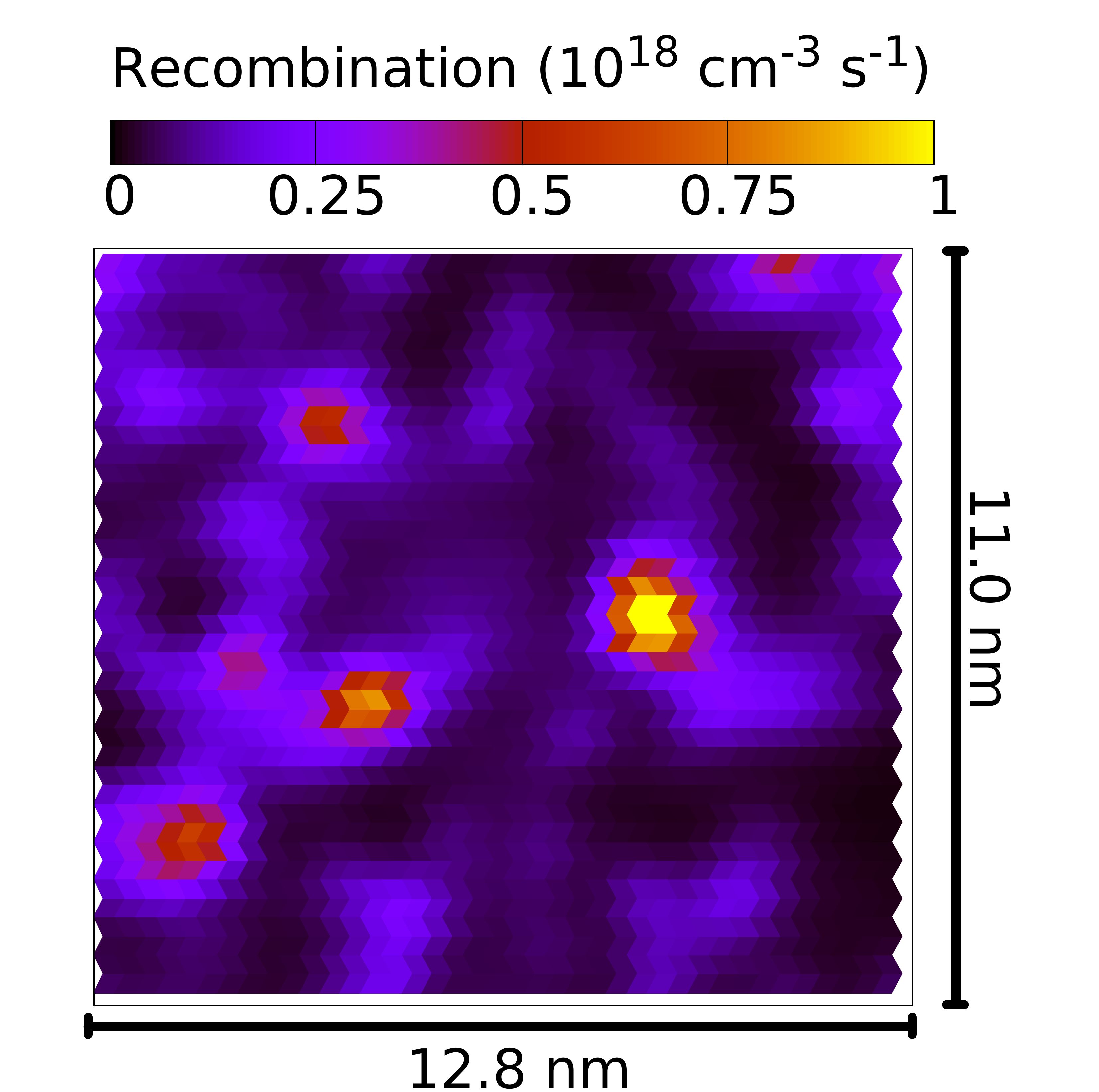} \\
    \hline
    \end{tabular}
    \caption{Profile of (a) valence band edge energy, (b) conduction band edge energy, and (c) radiative recombination rate in the growth plane ($c$-plane) of an In$_{0.1}$Ga$_{0.9}$N quantum well; the current density is $50\text{ A/cm}^2$ in all depicted figures. The slice displayed is the through the center well. The data are shown in all cases on a linear scale.}    
    \label{fig:BEs_2D}
\end{figure*}


To study the carrier distribution in MQW systems using the simulation settings (i) and (ii), we follow again the experimental approach e.g. presented in Ref.~\cite{BGaPhysStatSol_2011} and
the deep QW is moved from the $n$-side (position 1 ($n$-side) in Fig.~\ref{fig:schematic_of_QWs}) to the $p$-side (position 3 ($p$-side) in Fig.~\ref{fig:schematic_of_QWs}). In the case of the random alloy structures, the same microscopic configuration is kept for each well and only the ordering is changed. 

For each of these systems the ratio of radiative recombination from the shallow wells to the deep well is calculated using:

\begin{equation}
    \varrho = \frac{\mathcal{R}^{RAD}_{\Omega_S}}
    {\mathcal{R}^{RAD}_{\Omega_D}}
    \label{eq:RecombRatio}
\end{equation}

where 
\begin{equation}
    \mathcal{R}_{\Omega_i}^{RAD} = \int_{\Omega_i} R^{RAD}(\mathbf{r})dV\,\, ,
\end{equation} 
is the total radiative recombination from the region $\Omega_i$. 
Here, $\Omega_D$ is the region containing the deep QW, $\Omega_S$ is the region containing the shallow wells (as there are two shallow QWs this is the union of the two shallow QW regions). The radiative recombination rate at position $\mathbf{r}$, $R_{RAD}(\mathbf{r})$, is discussed in further detail in section~\ref{sec:TheoryDD}.
Since we are studying a system with three QWs, an even distribution of carriers across the MQWs would result in a ratio of $\varrho = 2$. 
Previous experimental work on a similar system by Galler \textit{et. al.}~\cite{BGaPhysStatSol_2011} found that \emph{$\varrho$ was small} (i.e. emission is dominated by the deep QW) \emph{only when the deep well was closest to the $p$-doped side of the MQW system} (thus position 3 ($p$-side) in Fig.\ref{fig:schematic_of_QWs}). The authors conclude that holes are responsible for this behavior, and argue that they are mainly found in the $p$-side QW and not in wells further away from the $p$-side. As a consequence, the overall emission from the (In,Ga)N/GaN MQW system is dominated by the emission from this well closest to the $p$-doped region. In line with Ref.~\cite{BGaPhysStatSol_2011}, we calculate $\varrho$ at a current density of 50 A/cm$^2$, which allows us to compare the here predicted trends with the trends found in the experimental studies. The theoretical framework employed to gain insight into $\varrho$ is discussed in the following section.

\section{Theoretical framework}\label{sec:Theory}
\label{sec:theory}

In this section, we introduce the underlying (microscopic) theory of our multiscale simulations. We start in Section~\ref{sec:TheoryTB} with the electronic structure model, an atomistic tight-binding (TB) model, and discuss the drift-diffusion approach in Section~\ref{sec:TheoryDD}. Since all these ingredients have been discussed in detail in Refs.~\cite{MiOD2021_JAP,OdFaOQE2022}, we here give only a brief summary.

\subsection{Tight-binding energy landscape}\label{sec:TheoryTB}

In order to model the electronic structure of the above described (In,Ga)N MQW systems on an atomistic level, we employ a nearest-neighbour, $sp^3$ TB model~\cite{ScCa2015}. Here, local strain and polarization effects are included using a valence force field model and local polarization theory, respectively~\cite{CaSc2013local}. 

To connect the TB model to a DD solver we proceed as follows. Firstly, the TB model is used to extract a potential energy landscape describing the MQW region of the device using an atomistic framework. To do so,  at each atomic site in the three dimensional (3-D) supercell, a local TB Hamiltonian is constructed from the full TB Hamiltonian~\cite{ChODo2021}. 
Subsequently, only the local TB Hamiltonian is diagonalized, yielding the conduction and valence band edge energy at each lattice site. These band edge energies include now already effects arising from alloy fluctuations and connected fluctuations in strain and built-in polarization field. The obtained 3-D confining energy landscape, after employing a Gaussian softening, forms the basis for our DD calculations.
In previous studies we have investigated and discussed in detail the influence of the Gaussian softening on transport calculations for electrons and holes~\cite{MiOD2021_JAP,OdFaOQE2022}. Here, we choose a Gaussian broadening on the order of the GaN lattice constant, $\sigma_{c,v} = a^\text{GaN} = 0.3189$ nm, in all calculations.
This value is large enough to average over a number of neighboring sites, while also small enough to retain fluctuations in the energy landscape. 

To obtain an accurate description of carrier transport in (In,Ga)N-based LEDs, the DD equations, which will be discussed below, are often coupled with solving the Schr\"odinger equation to account for quantum corrections.
Such a Schr\"odinger-Poisson solver is widely available in commercial software packages. However, in these packages it is largely restricted to 1-D simulations, since the extension to a 3-D system is computationally basically unfeasible. Instead of solving the large eigenvalue problem connected to evaluating the Schr\"odinger equation, we have implemented quantum corrections via localization landscape theory (LLT)~\cite{FiPi2017}; this approach is numerically much more efficient and gives results similar to a full self-consistent Schr\"odinger-Poisson solver, as we will also discuss below. 
From LLT we extract an effective confining potential for the conduction and valence band edge starting from the TB energy landscape. 

An example of the resulting quantum corrected energy landscape is given in Fig.~\ref{fig:BEs_2D}~(a) and (b). Here, in-plane band edge profiles for a single atomic plane through an In$_{0.1}$Ga$_{0.9}$N QW, after LLT has been applied, are shown. 
As Fig.~\ref{fig:BEs_2D} (a) reveals, the fluctuations in the valence band edge energy due to alloy fluctuations are of the order of 100 meV. In combination with the high effective hole mass, these fluctuations are large enough to give rise to strong carrier localization effects as seen in other studies already~\cite{WaGo2011,ScCa2015,DiPeJJAP2019}.
We therefore expect that, especially for holes, the inclusion of random alloy fluctuations in the simulation will impact the carrier distribution. Consequently  recombination rates are also expected to be noticeably influenced.

The variation in the conduction band edge energy is significantly smaller (order of 30 meV), as can be seen in Figure~\ref{fig:BEs_2D} (b). Since the effective electron mass is much lower in comparison with the holes, electron wave functions are less strongly perturbed by alloy fluctuations.
The impact that these fluctuations in the band edge energies have on the radiative recombination is also seen in Fig.~\ref{fig:BEs_2D}~(c); the radiative recombination is calculated with \texttt{ddfermi} as will be described in section~\ref{sec:TheoryDD}. The correlation between the valence band edge maxima and regions of high radiative recombination can be clearly identified; similar spatial profiles can be seen for non-radiative (Auger) recombination (not shown). 


In order to highlight the impact of random alloy fluctuations on carrier transport and the distribution of carriers across a MQW system, we compare our atomistic calculations with the outcome of a VCA. In the latter case a homogeneous effective crystal is constructed where material properties are chosen to be interpolated properties of the binaries InN and GaN within the QW region. Here a linear, composition weighted interpolation scheme is employed. A bandgap bowing of $-2.0$ eV is used, consistent with the underlying atomistic TB model~\cite{CaSc2013local}. The VCA description, without any Gaussian broadening, is similar to commercially available packages. However, and in contrast to commercial software packages, quantum corrections via LLT can also be taken into account in our VCA simulations, following the approach used for the random alloy case.


\subsection{Device simulation}\label{sec:TheoryDD}
Having outlined above the generation of the energy landscape of the active region, e.g. the (In,Ga)N/GaN MQW system, a full device mesh, including the $n$- and $p$-doped regions, needs to be constructed on which the DD equations are solved. To achieve this we proceed as follows and
divide the device mesh into two regions: an atomistic and a macroscopic one. 
The atomistic region is used to describe the MQW region and has as many grid points as atoms in the system. These points contain information about the conduction and valence band edge energies calculated from TB, as discussed above. 

In order to capture the effects of carrier localization in the calculations, the in-plane dimensions of our 3-D simulation cell should be larger than the localization length of the holes, given that electrons are less strongly affected by alloy fluctuations~\cite{TaCaRSC2016}. In our atomistic calculations we use a system with in-plane dimensions of $12.8 \times 11.0\text{ nm}^2$. This is large enough to see the effects of hole localization as the in-plane hole localization length for  
In$_{0.1}$Ga$_{0.9}$N QWs is of the order of 1 nm~\cite{TaDa2020}. The in-plane dimensions can be seen in Fig.~\ref{fig:BEs_2D}~(a)~and~(b) where the in-plane valence and conduction band edges of an In$_{0.1}$Ga$_{0.9}$N QW are shown. In case of the VCA, given that there are no variations in material properties (band edge energies) within the growth plane ($c$-plane), a much smaller in-plane area is sufficient ($1.3 \times 1.1\text{ nm}^2$), which reduces the numerical effort.

LLT is solved on this (finite element) mesh using a finite element method (FEM). 
The DD calculations are carried out employing a Voronoi finite volume method (FVM)~\cite{Farrell2017}. Therefore, the generated FEM mesh must be transferred to an appropriate FVM mesh.
Every point from the FEM mesh is included on the FVM mesh, as well as extra points required to produce a boundary-conforming Delaunay tetrahedral mesh; conduction and valence band data are then interpolated onto these additional nodes.
This mesh is then embedded within a macroscopic device mesh which contains information about the $p$- and $n$-doped regions. In our atomistic transport studies here, we focus on systems without an (Al,Ga)N electron blocking layer (EBL). In principle, an atomistic mesh resolution would be required for the EBL, given the alloy fluctuations in (Al,Ga)N. However, and as we will discuss below, an (Al,Ga)N EBL is of secondary importance for the questions targeted in the present study. Therefore, outside active MQW region of the system, pure GaN is assumed. Thus, the conduction and valence band edge values are position independent (except for changes due an applied bias). The absence of strongly fluctuating band edges in the macroscopic mesh region allows us to use a sparse mesh and scale the simulation to a full device.
The mesh is created using \texttt{TetGen}~\cite{Si15ACM} and the interpolation is handled via \texttt{WIAS-pdelib}~\cite{pdelib}.  More details on the mesh generation can be found in Ref.~\cite{MiOD2021_JAP}.

Equipped with knowledge about the mesh generation, we turn now to the DD simulations. To do so we build on the van Roosbroeck system of equations~\cite{VanRoosbroeck1950}:

\begin{subequations}
    \begin{equation}
        -\nabla \cdot (\varepsilon_s(\mathbf{r})\nabla\psi(\mathbf{r})) = q(p(\mathbf{r})-n(\mathbf{r})+C(\mathbf{r}))\,\, ,
        \label{eq:Poisson}
    \end{equation}
    \begin{equation}
        \nabla\cdot \mathbf{j}_n(\mathbf{r}) = qR(\mathbf{r})\,\, ,
        \label{eq:current_continuity_elec}
    \end{equation}
    \begin{equation}
        \nabla\cdot \mathbf{j}_p(\mathbf{r}) = -qR(\mathbf{r})\,\, ,
        \label{eq:current_continuity_hole}
    \end{equation}
    \begin{equation}
        \mathbf{j}_n(\mathbf{r}) = -q \mu_n n(\mathbf{r}) \nabla\varphi_n(\mathbf{r})\,\, ,
        \label{eq:current_elec}
    \end{equation}
    \begin{equation}
        \mathbf{j}_p(\mathbf{r}) = -q \mu_p p(\mathbf{r}) \nabla\varphi_p(\mathbf{r})\,\, .
        \label{eq:current_hole}
    \end{equation}
    \label{eq:VanRoosbroeck}
\end{subequations}


In the above equations, $q$ is the elementary charge, \mbox{$\varepsilon_s(\mathbf{r}) = \varepsilon_0\varepsilon_r (\mathbf{r})$} is the dielectric permittivity, $\psi(\mathbf{r})$ is the device electrostatic potential, $p(\mathbf{r})$ and $n(\mathbf{r})$ are the hole and electron densities, $C(\mathbf{r}) = N_D^+(\mathbf{r}) - N_A^+(\mathbf{r})$ is the net activated dopant density, $\mathbf{j}_n(\mathbf{r})$, $\mathbf{j}_p(\mathbf{r})$, $\varphi_n(\mathbf{r})$ and $\varphi_p(\mathbf{r})$ are the electron and hole current densities and the respective quasi-Fermi potentials. The total recombination rate is denoted by $R$.

The carrier densities are related to the band edge energies, $E^\text{dd}_c(\mathbf{r})$ and $E^\text{dd}_v(\mathbf{r})$, and quasi-Fermi potentials, $\varphi_n(\mathbf{r})$ and $\varphi_p(\mathbf{r})$, via the state equations~\cite{Farrell2017}

\begin{subequations}
    \begin{equation}
        n(\mathbf{r}) = N_c \mathcal{F}\Bigg( \frac{q(\psi(\mathbf{r})-\varphi_n(\mathbf{r}))-E_c^{dd}(\mathbf{r})}{k_BT}\Bigg)\, \, , 
    \end{equation}
    \begin{equation}
        p(\mathbf{r}) = N_v \mathcal{F}\Bigg( \frac{E_v^{dd}(\mathbf{r}) - q(\psi(\mathbf{r})-\varphi_p(\mathbf{r}))}{k_BT} \Bigg)\,\, .
    \end{equation}
\end{subequations}

Here, $N_c$ and $N_v$ are the effective density of states for the conduction and valence band, respectively. For the distribution function, $\mathcal{F}$, we use  Fermi-Dirac statistics, and $k_B$ is the Boltzmann constant. A temperature of $T = 300$ K has been used in all calculations. It is to note that the valence band, $E_v^{dd}(\mathbf{r})$, and conduction band edge energy, $E_c^{dd}(\mathbf{r})$, can be described either by a VCA, a VCA including quantum corrections via LLT, or an atomistic random alloy calculation including LLT-based quantum corrections.

The total recombination rate in Eqs.~(\ref{eq:current_continuity_elec}) and~(\ref{eq:current_continuity_hole}), $R$, is calculated using the ABC model~\cite{KaOQE2015,PiPhysStatSol2010}. Here, $R$ is the sum of (defect related) Shockley-Read-Hall, $R^{SRH}$, radiative, $R^{RAD}$ and (non-radiative) Auger recombination rate, $R^{AUG}$.
The SRH rate is obtained from:
    \begin{equation}
        R^{SRH}(\mathbf{r}) = \frac{r(n,p)}{\tau_p\big(n(\mathbf{r}) + n_i(\mathbf{r}) \big) + \tau_n\big(p(\mathbf{r})+n_i(\mathbf{r}) \big)}\,\, ,
        \label{eq:R_SRH}
    \end{equation}
the radiative part via
    \begin{equation}
        R^{RAD}(\mathbf{r}) = B_0 r(n,p)\,\, ,
        \label{eq:R_RAD}
    \end{equation}
    and the Auger rate is calculated as
    \begin{equation}
        R^{AUG}(\mathbf{r}) = ( C_n n(\mathbf{r}) + C_p p(\mathbf{r}) ) r(n,p)\,\, .
        \label{eq:R_AUG}
    \end{equation}
In Eqs.~(\ref{eq:R_SRH}) to~(\ref{eq:R_AUG}) 
$$r(n,p) = n(\mathbf{r})p(\mathbf{r}) - n_i^2(\mathbf{r})$$
and
$$
n_i^2(\mathbf{r}) = n(\mathbf{r})p(\mathbf{r})\exp\bigg(\frac{q\varphi_n - q\varphi_p}{k_BT}\bigg).
$$

The above equations require further input, namely the radiative recombination coefficient $B_0$, the Auger recombination coefficients $C_p$ and $C_n$ as well as the the SRH lifetimes $\tau_p$ and $\tau_n$. All these parameters will in principle carry a composition dependence~\cite{McKiPRB2022,McTa2020,KiQiNJOP2013}. Furthermore, $B_0$, $C_p$ and $C_n$ will also be carrier density dependent~\cite{DaYoPRA2019,DaNaECS2019,JoTeAPL2017}.
We follow here the widely made assumption that these coefficients are constant across the InGaN MQW region~\cite{LiPi2017,LyLhPhysRevApplied2022}. In the following we take a weighted average of parameters calculated in Ref.~\cite{McKiPRB2022} for an electron and hole density of $3.8\times 10^{18}$ cm$^{-3}$, which is a good approximation for the average carrier densities in the QWs at a current density of 50 A/cm$^2$. 
As our active region consists of two In$_{0.1}$Ga$_{0.9}$N QWs and one In$_{0.125}$Ga$_{0.875}$N QW we evaluate the different recombination coefficients as follows:
\begin{equation}
    R_i^{\text{eff}} = \frac{2\times(R_i^{10\%}) + 0.5\times(R_i^{15\%} + R_i^{10\%})}{3}\,\, .
\end{equation}
Here, $R_i \in \{B_0,C_n,C_p\}$ are the radiative recombination, electron-electron-hole and hole-hole-electron Auger recombination coefficients, respectively. As there are no values for an In$_{0.125}$Ga$_{0.875}$N QW in Ref.~\cite{McKiPRB2022}, a linear average of the coefficients in In$_{0.1}$Ga$_{0.9}$N and In$_{0.15}$Ga$_{0.85}$N wells has been used. A summary of the material parameters employed in all simulations is given in Table~\ref{tab:MatParams}.



\begin{table}[t]
\caption{Material parameters used in the different regions of the simulation supercell. Parameters denoted with $\dagger$ are taken from~\cite{LiPi2017}; parameters denoted with  $\ddagger$ are derived from~\cite{McKiPRB2022} as described in the main text.}
    \label{tab:MatParams}
    \centering
    \begin{tabular}{|lr|c||r|r|r|}
    \hline
    \multicolumn{3}{|c||}{Parameter} & \multicolumn{3}{c|}{Value in each region} \\
    \hline
        \multicolumn{1}{|c}{\textbf{Name}}      & &
        \multicolumn{1}{|c||}{\textbf{Units}}    & 
        \multicolumn{1}{c}{\textbf{$p$-GaN}}   & 
        \multicolumn{1}{|c}{\textbf{$i$-InGaN}} & 
        \multicolumn{1}{|c|}{\textbf{$n$-GaN}}   \\
    \hline
    Doping &&cm$^{-3}$ &  $5\times 10^{18}$ & $1\times 10^{16}$ & $5\times 10^{18}$ \\
    $\mu_h$&$^\dag$ &cm$^2$/Vs & 5 & 10 & 23 \\
    $\mu_e$&$^\dag$ &cm$^2$/Vs & 32 & 300 & 200 \\
    $\tau_p$&$^\dag$ &s & $10$ & $1\times10^{-7}$ & $7\times10^{-10}$ \\
    $\tau_n$&$^\dag$ &s & $6\times10^{-10}$ & $1\times10^{-7}$ & $10$ \\
    $B_0$&$^\ddagger$ &cm$^3$/s& $2.8\times10^{-11}$ & $2.8\times10^{-11}$ & $2.8\times10^{-11}$ \\
    $C_p$&$^\ddagger$ &cm$^6$/s & $5.7\times10^{-30}$ & $5.7\times10^{-30}$ & $5.7\times10^{-30}$ \\
    $C_n$&$^\ddagger$ &cm$^6$/s & $1\times10^{-31}$ & $1\times10^{-31}$ & $1\times10^{-31}$ \\
    \hline
    \end{tabular}
\end{table}


The numerical approximation of the van Roosbroeck system is implemented (in 3-D) in \texttt{ddfermi}~\cite{ddfermi}. As already mentioned above, we employ the FVM and the current is discretized using the SEDAN (excess chemical potential) approach 
\cite{Yu1988,Cances2020,Abdel2020b}, which yields a thermodynamically consistent flux approximation in the sense of Ref.~\cite{Farrell2017}.

To simulate the devices under study, we also used the commercial software \NN{~\cite{nextnano}}, which relies on the simulation of a self-consistent Schr\"odinger-Poisson-DD system.
In this work we use \NN{ }to simulate the carrier transport in the above discussed MQW systems within a 1-D approximation. 
In \NN{ }we utilize the same parameter set as in the \texttt{ddfermi} simulations. Therefore, the obtained results can be directly compared to our 3-D VCA model. When including quantum corrections in \texttt{ddfermi}, LLT is used. In \NN{ } a self-consistent Schr\"odinger-Poisson-DD calculation is performed where a $\mathbf{k}\cdot\mathbf{p}$ Hamiltonian
is used to calculate eigenstates across the full simulation domain.
Following the \texttt{ddfermi} set up, in \NN{ }we employ also a 1-band model for the calculation of the electron and hole densities.

\section{Results}\label{sec:Results}

\begin{figure*}[t]
    \centering
    \begin{tabular}{|c|c|}
        \hline
        \large{\textbf{(a)} \NN{}} (1-D) & \large{\textbf{(b)} \texttt{ddfermi}} (3-D)\\
        \hline
        \text{ }\vspace{-5pt} & \text{ } \\
        \includegraphics[width=0.3\textwidth]{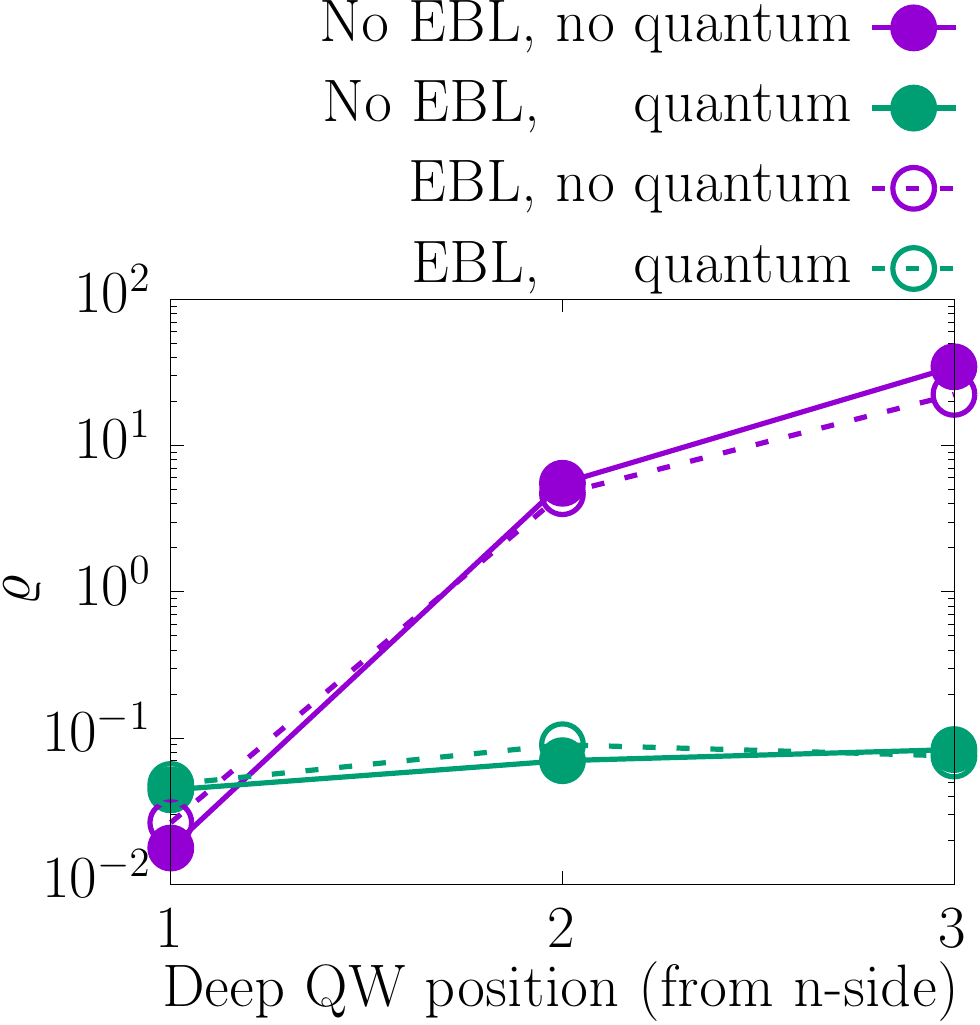} &
        \includegraphics[width=0.3\textwidth]{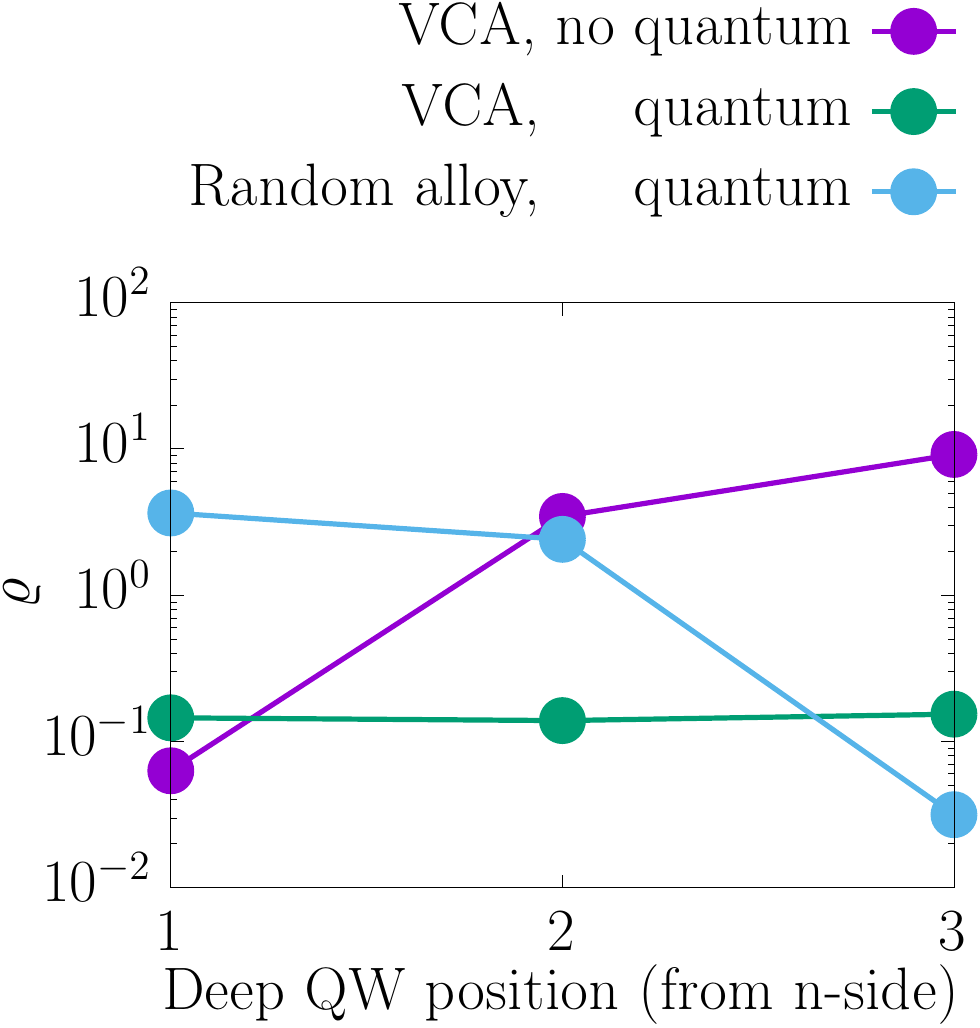} \\
        \hline
    \end{tabular}
    \caption{Ratio of radiative recombination $\varrho$, Eq.~(\ref{eq:RecombRatio}), from the shallow wells (In$_{0.1}$Ga$_{0.9}$N) to recombination from the deep well (In$_{0.1}$Ga$_{0.9}$N) calculated as a function of the position of the deep well in the multi-quantum well stack. Here $\varrho$ is evaluated using (a) \NN{ }excluding (purple) and including (green) quantum corrections via a self-consistent Schr\"odinger-Poisson-drift diffusion solver; results are shown when excluding (solid, filled circles) and including (dotted, open circles) an Al$_{0.15}$Ga$_{0.85}$N blocking layer, and (b) \texttt{ddfermi} excluding (purple), including (green) quantum corrections via localization landscape theory (LLT) using a virtual crystal approximation (VCA) and a random alloy calculation including LLT-based quantum corrections (blue); these calculations neglect the AlGaN blocking layer.}
    \label{fig:recombratio}
\end{figure*}

In this section we present the results of our study on the carrier distribution in the above described (In,Ga)N/GaN MQW systems. To understand the impact of the alloy microstructure on the carrier distribution, in Section~\ref{sec:ResultsEBL} we start with standard 1-D calculations building on the commercial software package \NN{~\cite{nextnano}}. We use this entirely continuum-based description of the QWs also to determine the impact (i) of an EBL and (ii) a self-consistent Schrodinger-Poisson-DD treatment on the transport properties. 
Moreover, and as already mentioned above, (ii) can also be used as a benchmark for our 3-D \texttt{ddfermi} solver. In  Section~\ref{sec:ResultsFluctuations} we then proceed to investigate the influence of random alloy fluctuations on the carrier distribution in the (In,Ga)N/GaN MQW stack. 

\subsection{Continuum-based simulations of the carrier transport in (In,Ga)N-based LEDs}\label{sec:ResultsEBL}

\begin{figure*}
    \centering
    \begin{tabular}{|c|c|c|}
    \hline
        \large \textbf{(a) VCA} \normalsize  & \large \textbf{(b) VCA + LLT} \normalsize  & \large \textbf{(c) Random alloy + LLT} \normalsize \\
    \hline
    \text{ } \vspace{-5pt} & & \\
        \includegraphics[width=0.3\textwidth]{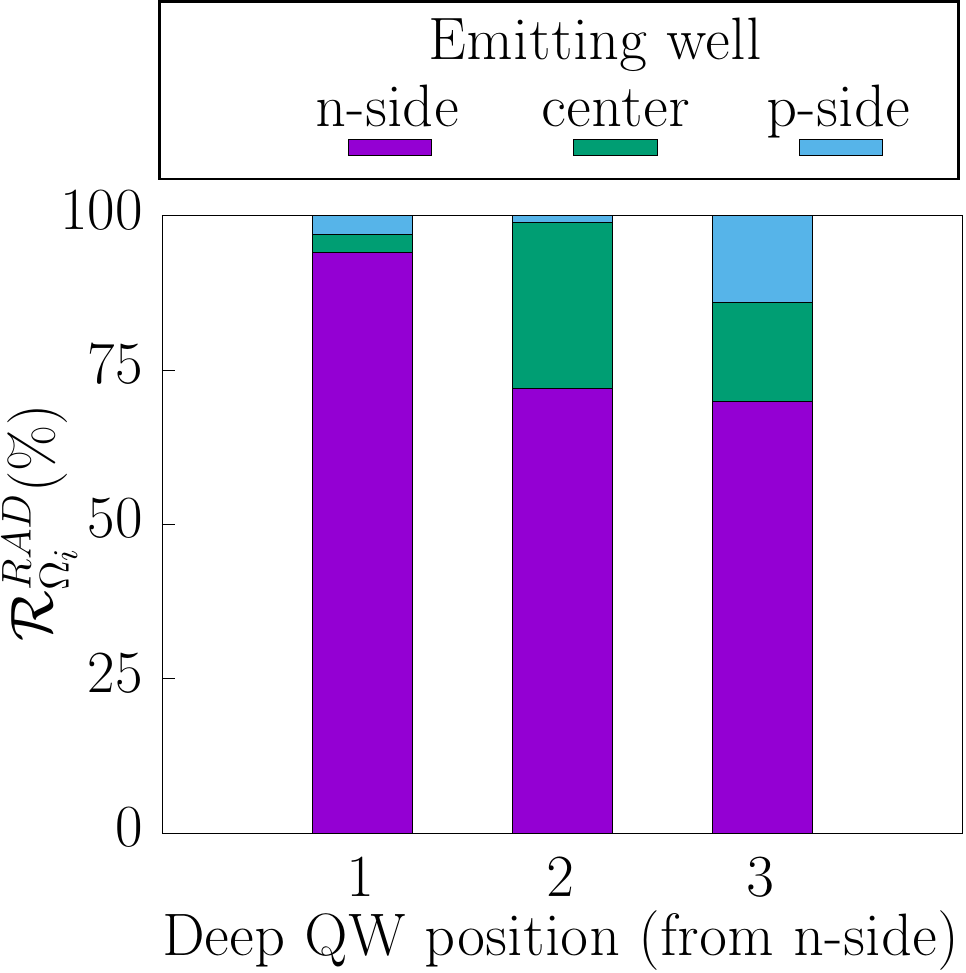} &
        \includegraphics[width=0.3\textwidth]{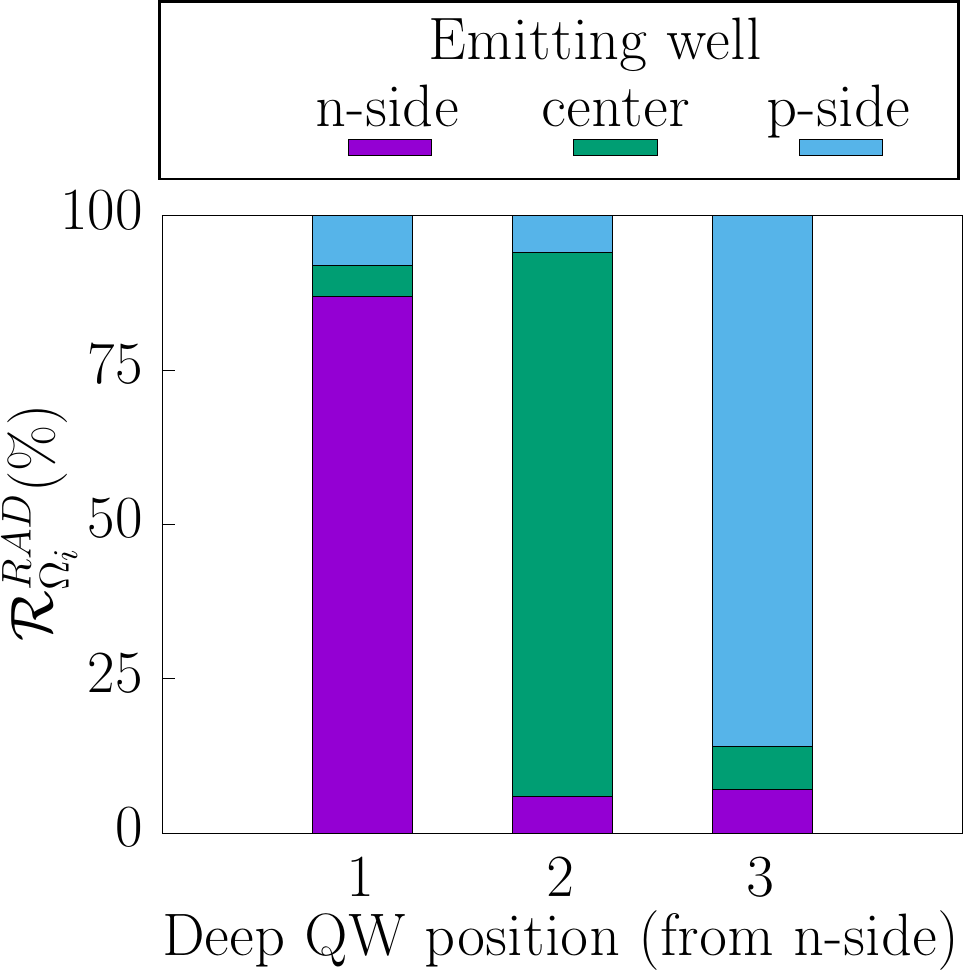} &
        \includegraphics[width=0.3\textwidth]{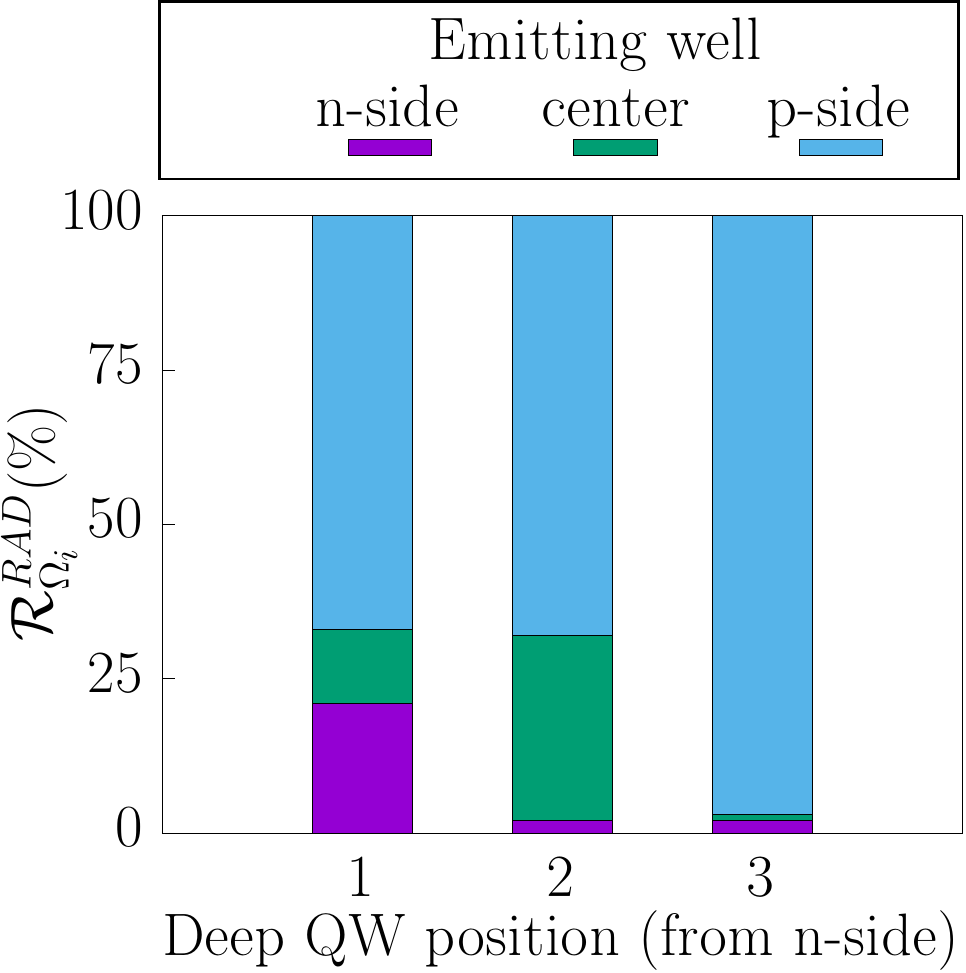}\\
    \hline
    \end{tabular}
    \caption{
    Contribution of each quantum well ($n$-side; centre; $p$-side) in the (In,Ga)N multi-quantum well system to the total radiative recombination $\mathcal{R}^{RAD}_{\Omega_i}$ for $i\in \{n\text{-side},\text{center},p\text{-side}\}$ as a percentage of the total radiative recombination from all 3 quantum wells for (a) virtual crystal approximation (VCA), (b) virtual crystal approximation with quantum corrections included via localization landscape theory (VCA + LLT) and (c) a random alloy calculation including localization landscape theory based quantum corrections (Random alloy + LLT). That data are shown as a function of the position of the deep quantum well ($x$-axis). Each bar contains the percentage recombination from the $n$-side quantum well (purple), the center quantum well (green) and the $p$-side quantum well (blue). Labelling is consistent with that introduced in Fig.~\ref{fig:schematic_of_QWs}. 
    }
    \label{fig:percentagecontribution}
\end{figure*}

\begin{figure*}[t]
    \centering
    \begin{tabular}{|c||c||c|}
        \hline
        \large \textbf{(a)} $n$-side \normalsize & \large \textbf{(b)} center \normalsize  & \large  \textbf{(c)} $p$-side \normalsize \\
        \hline
        \hline
        \multicolumn{3}{|c|}{\large \textbf{(i) VCA}} \\
        \hline
            \includegraphics[width=0.3\textwidth]{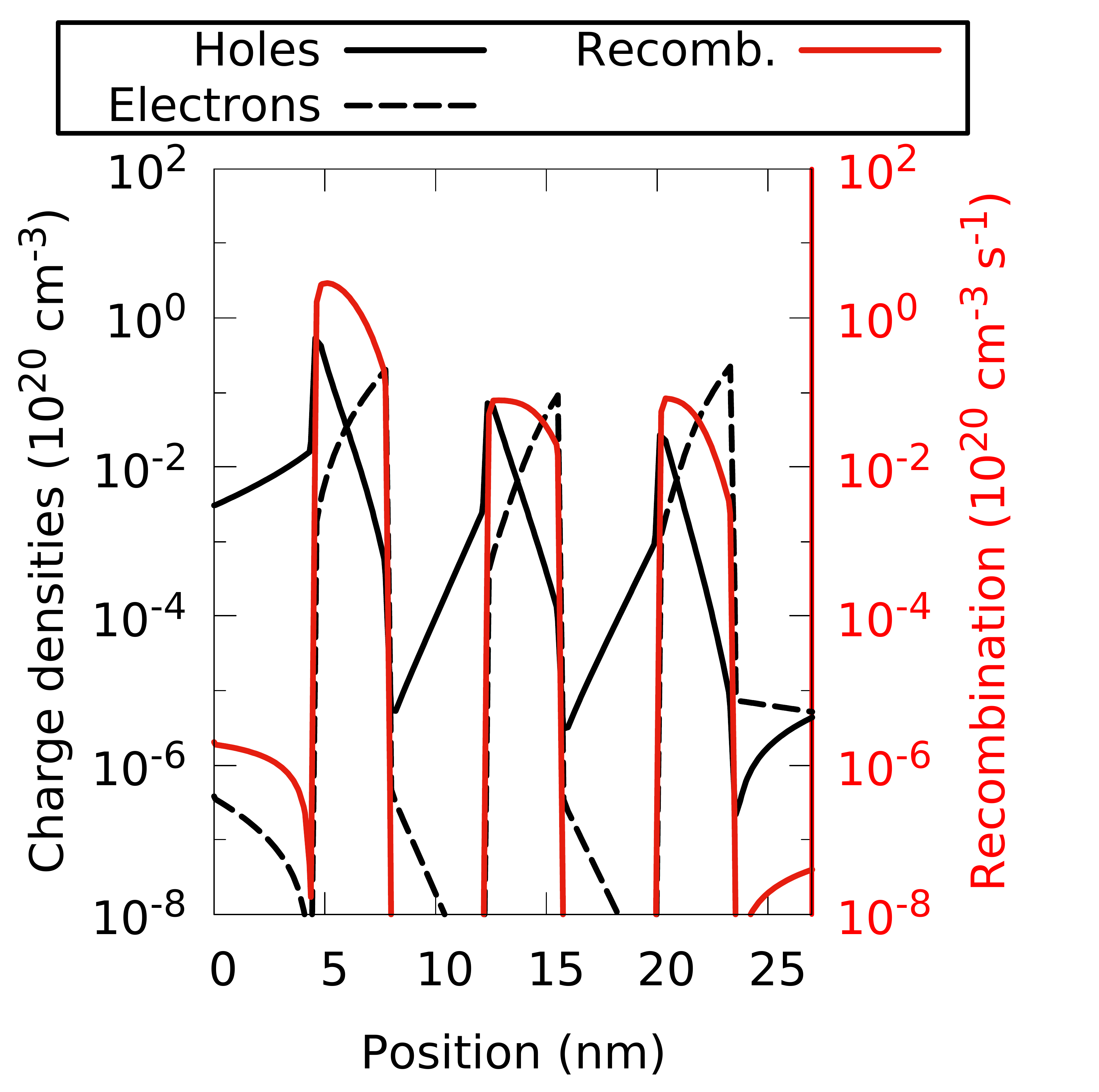} &
            \includegraphics[width=0.3\textwidth]{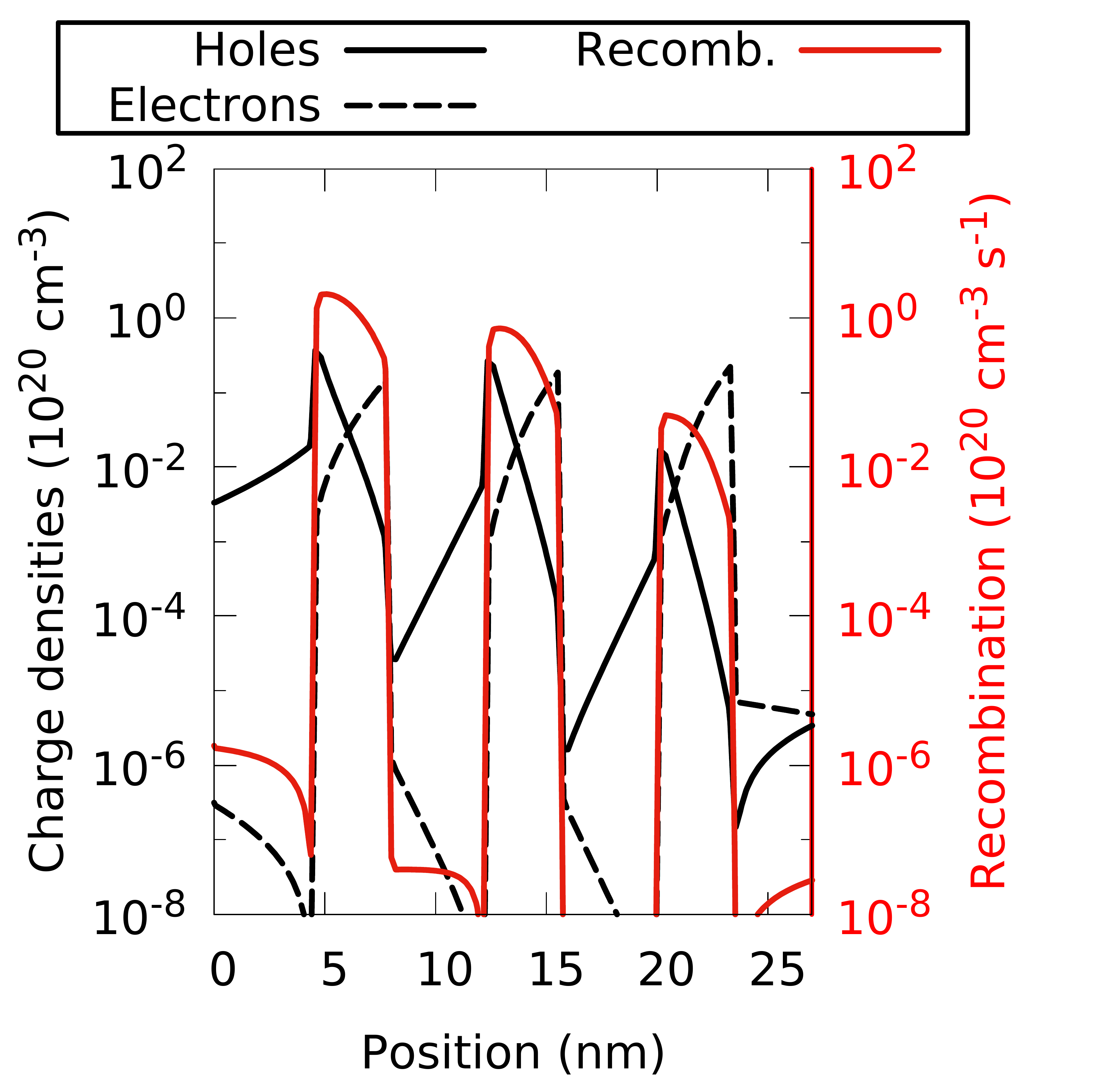} &
            \includegraphics[width=0.3\textwidth]{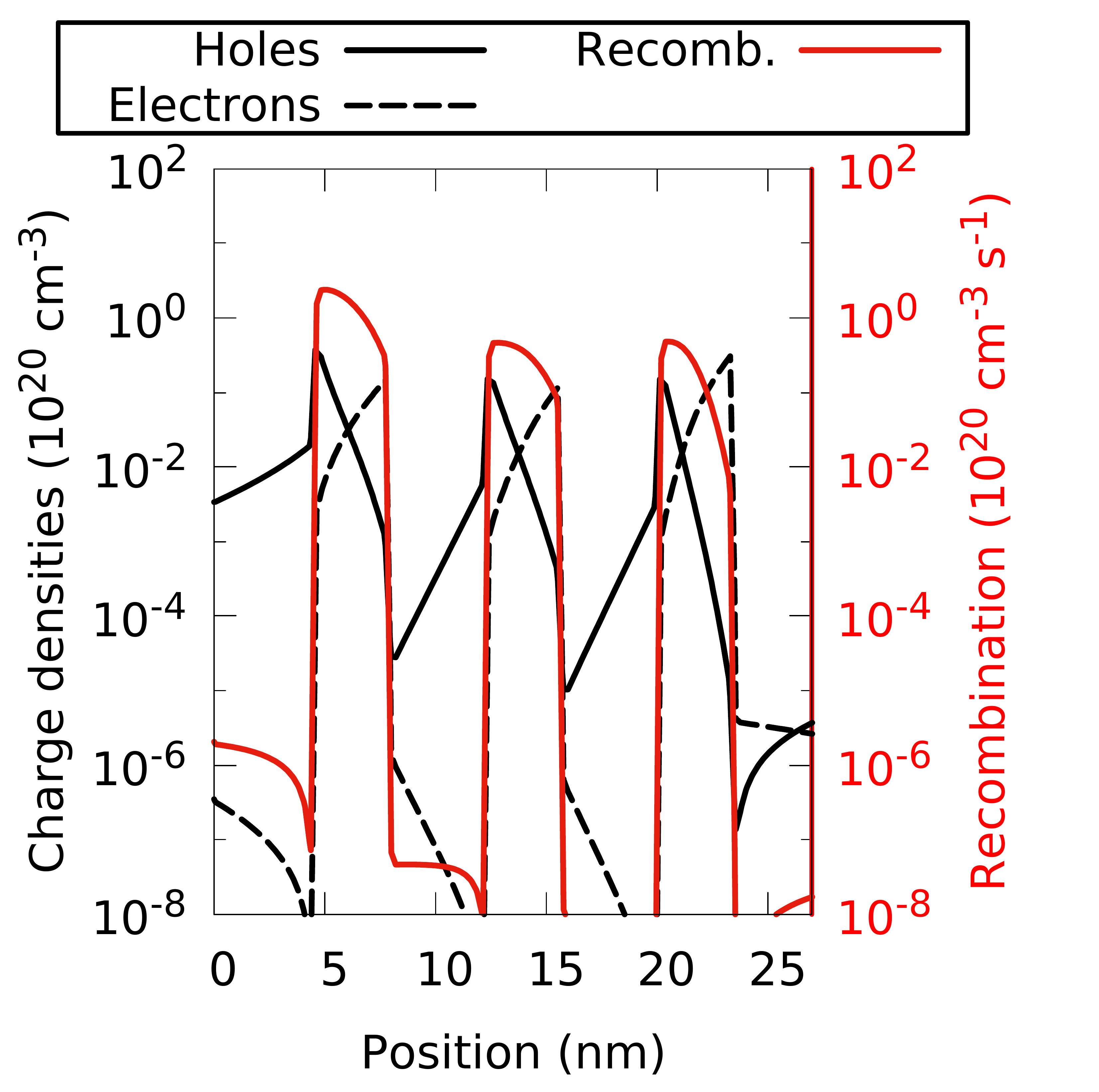} \\
        \hline
        \hline
        \multicolumn{3}{|c|}{\textbf{\large (ii) VCA + LLT}}\\
        \hline
            \includegraphics[width=0.3\textwidth]{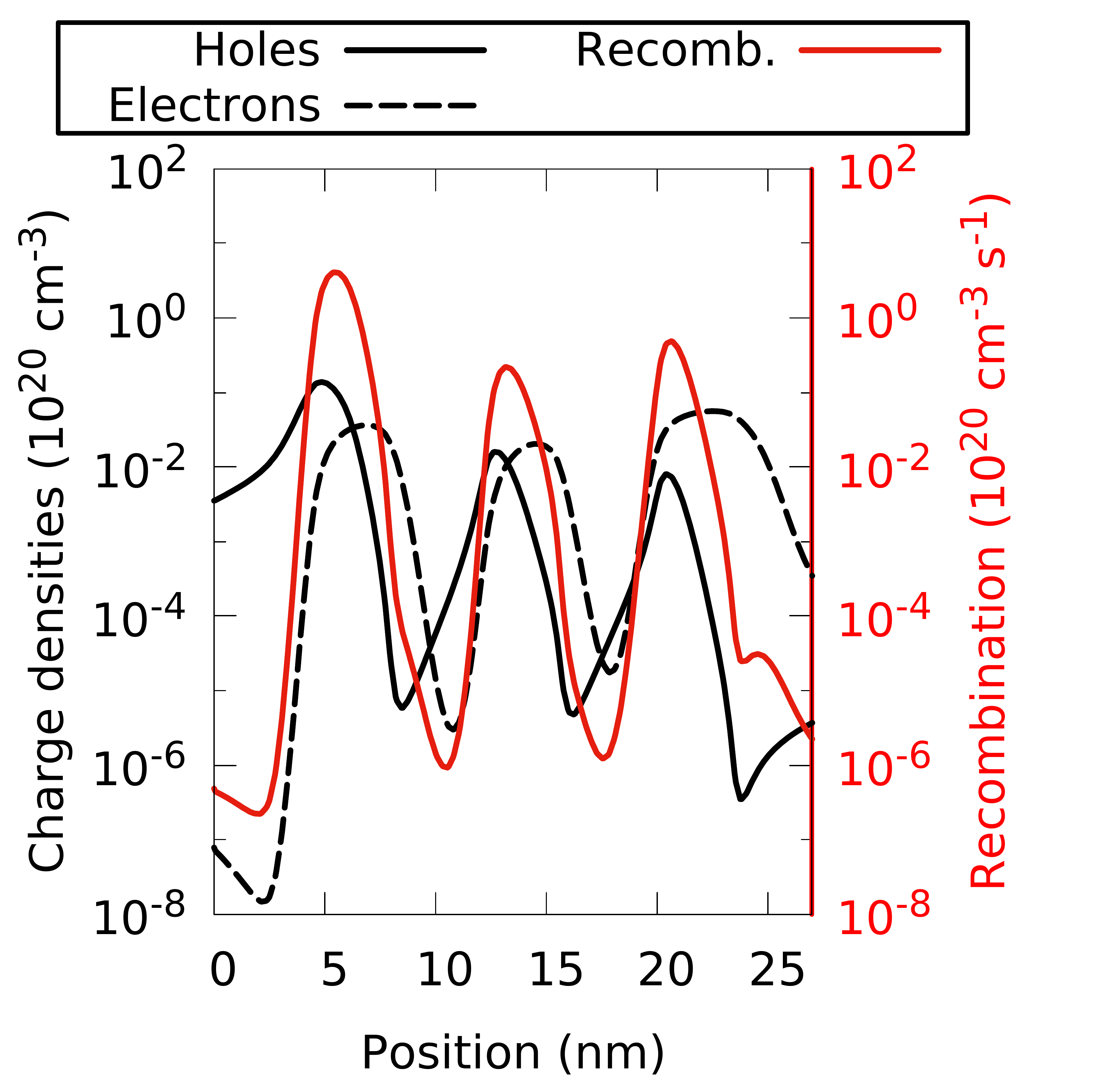} &
            \includegraphics[width=0.3\textwidth]{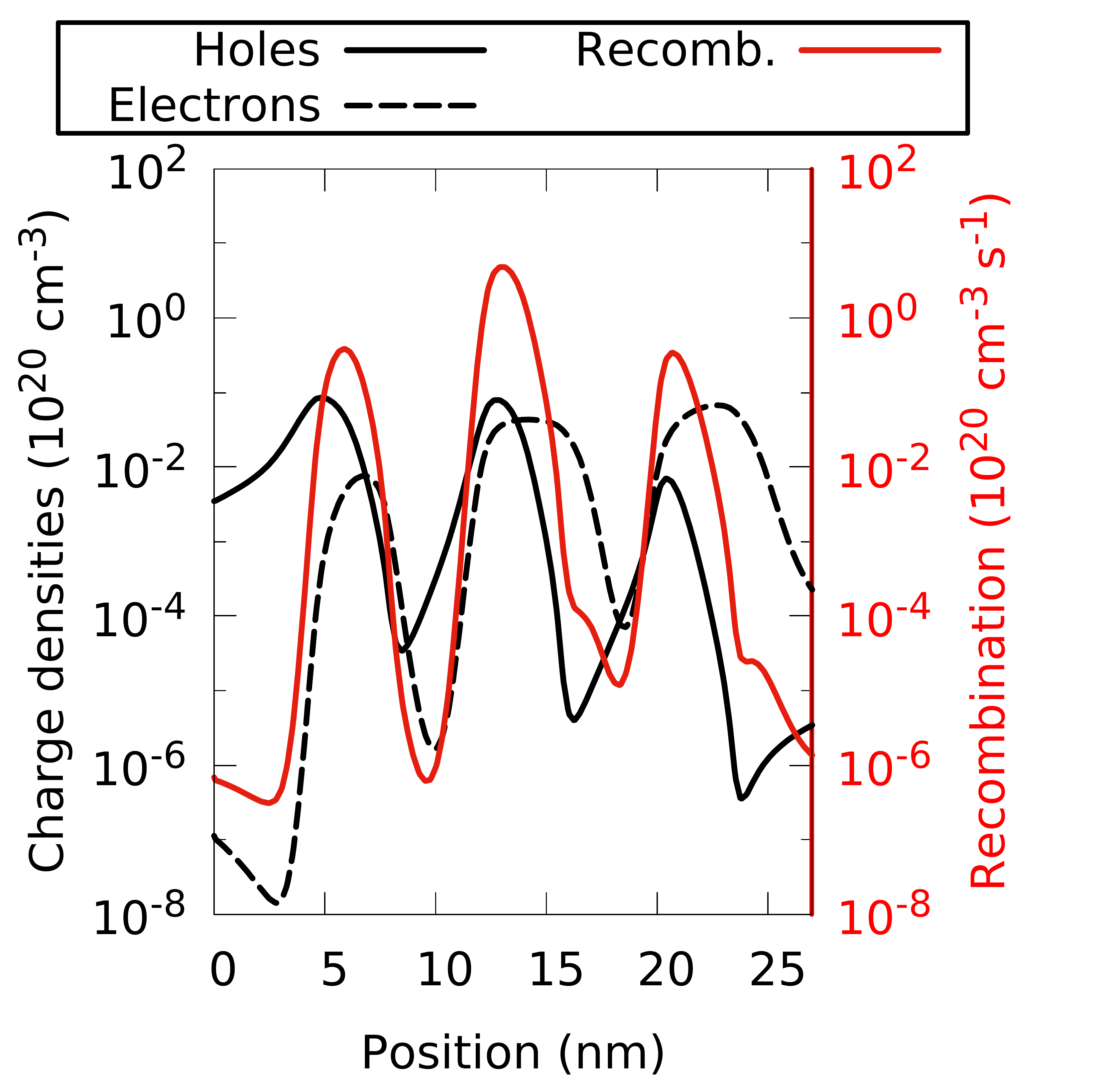} &
            \includegraphics[width=0.3\textwidth]{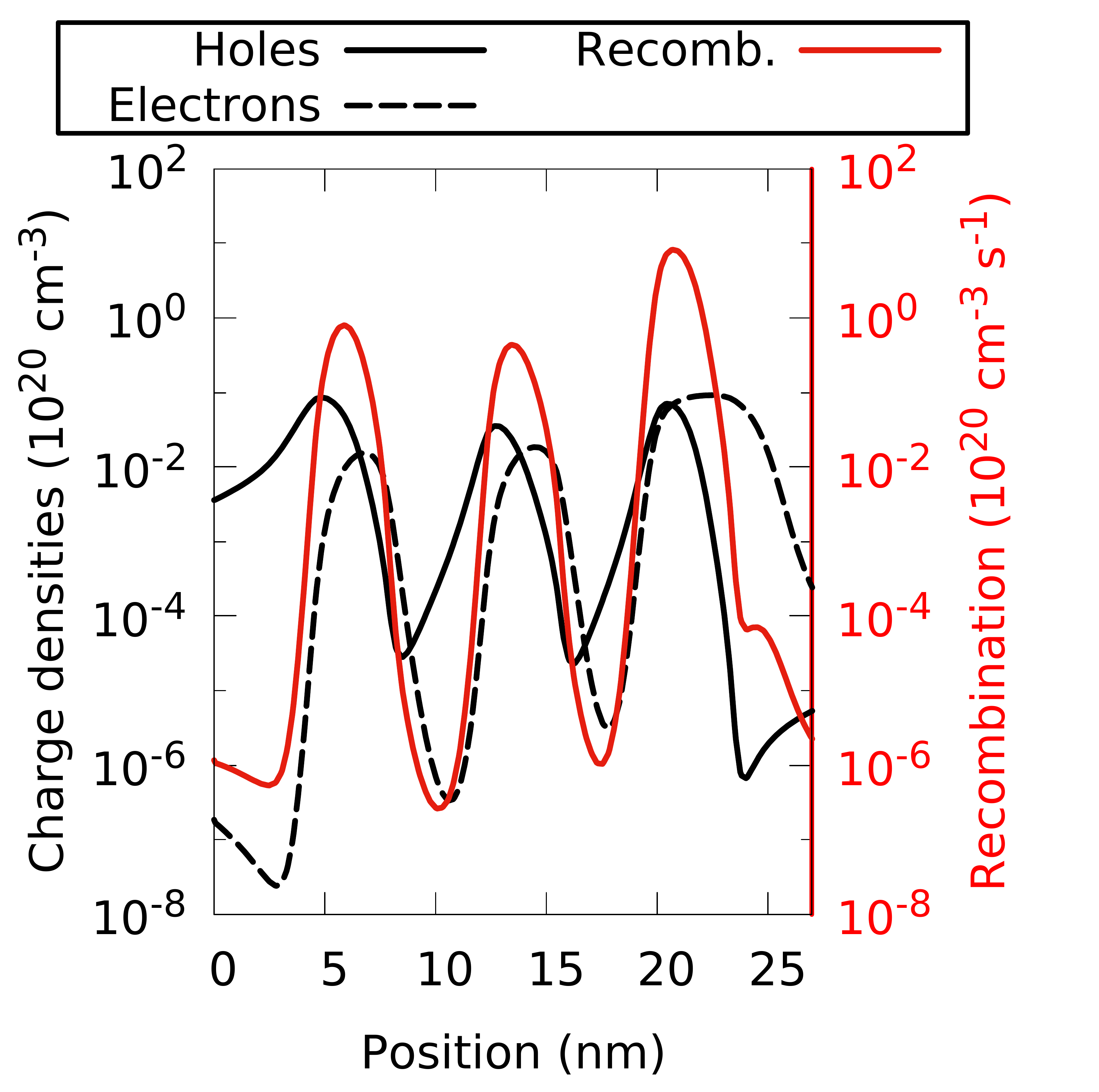} \\
        \hline
        \hline
        \multicolumn{3}{|c|}{\large \textbf{(iii) Random alloy + LLT}} \\
        \hline
            \includegraphics[width=0.3\textwidth]{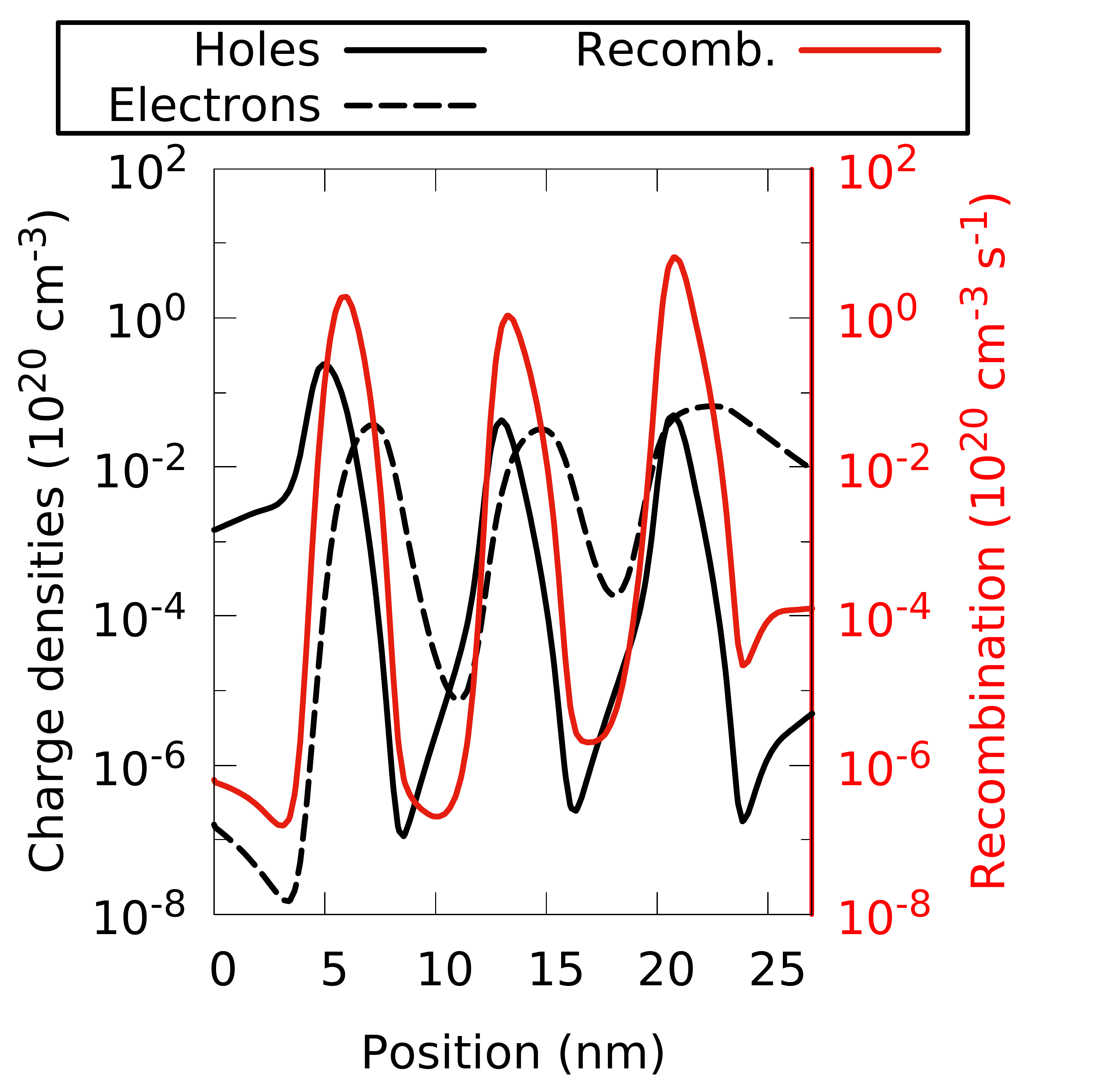} &
            \includegraphics[width=0.3\textwidth]{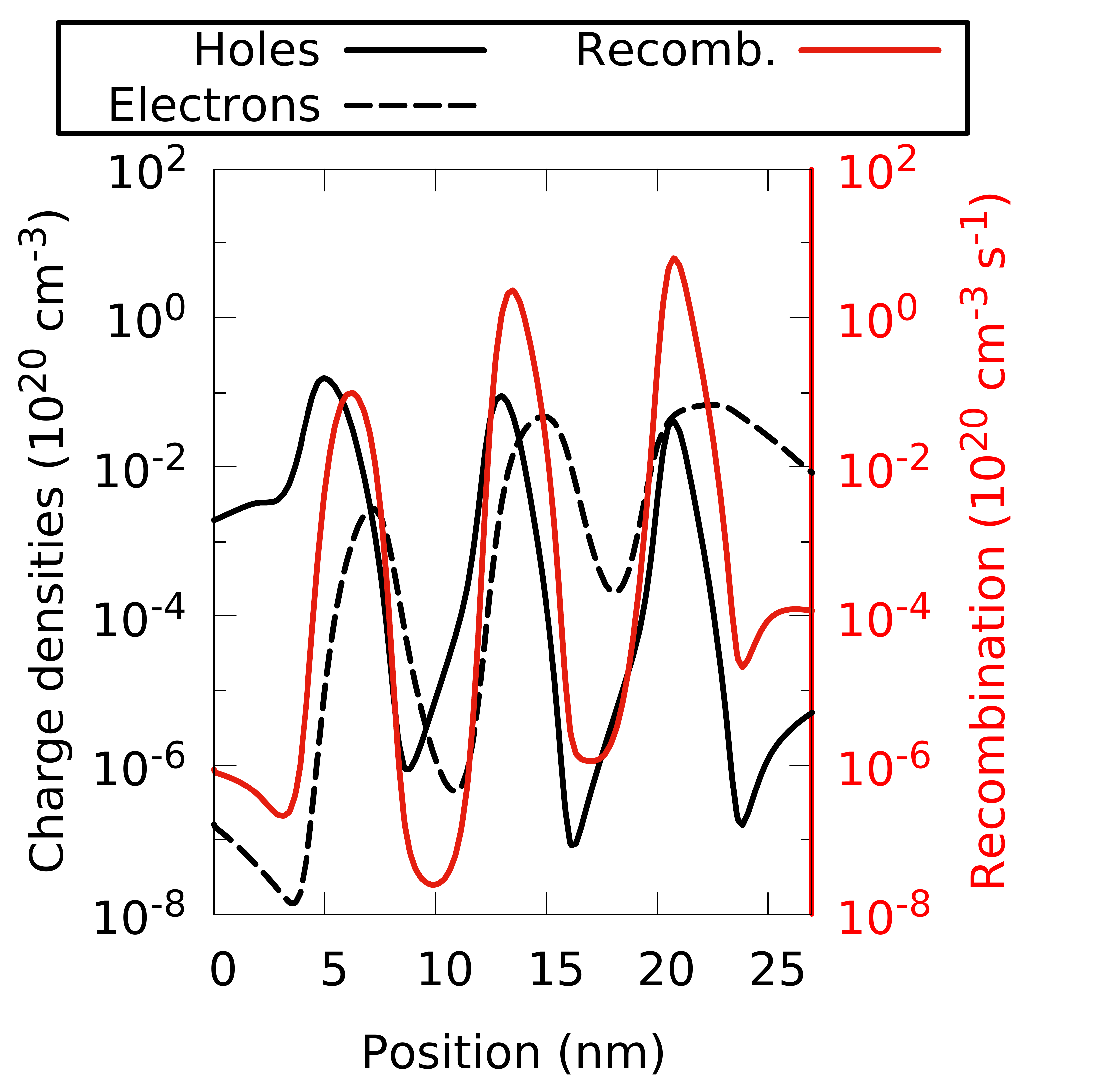} &
            \includegraphics[width=0.3\textwidth]{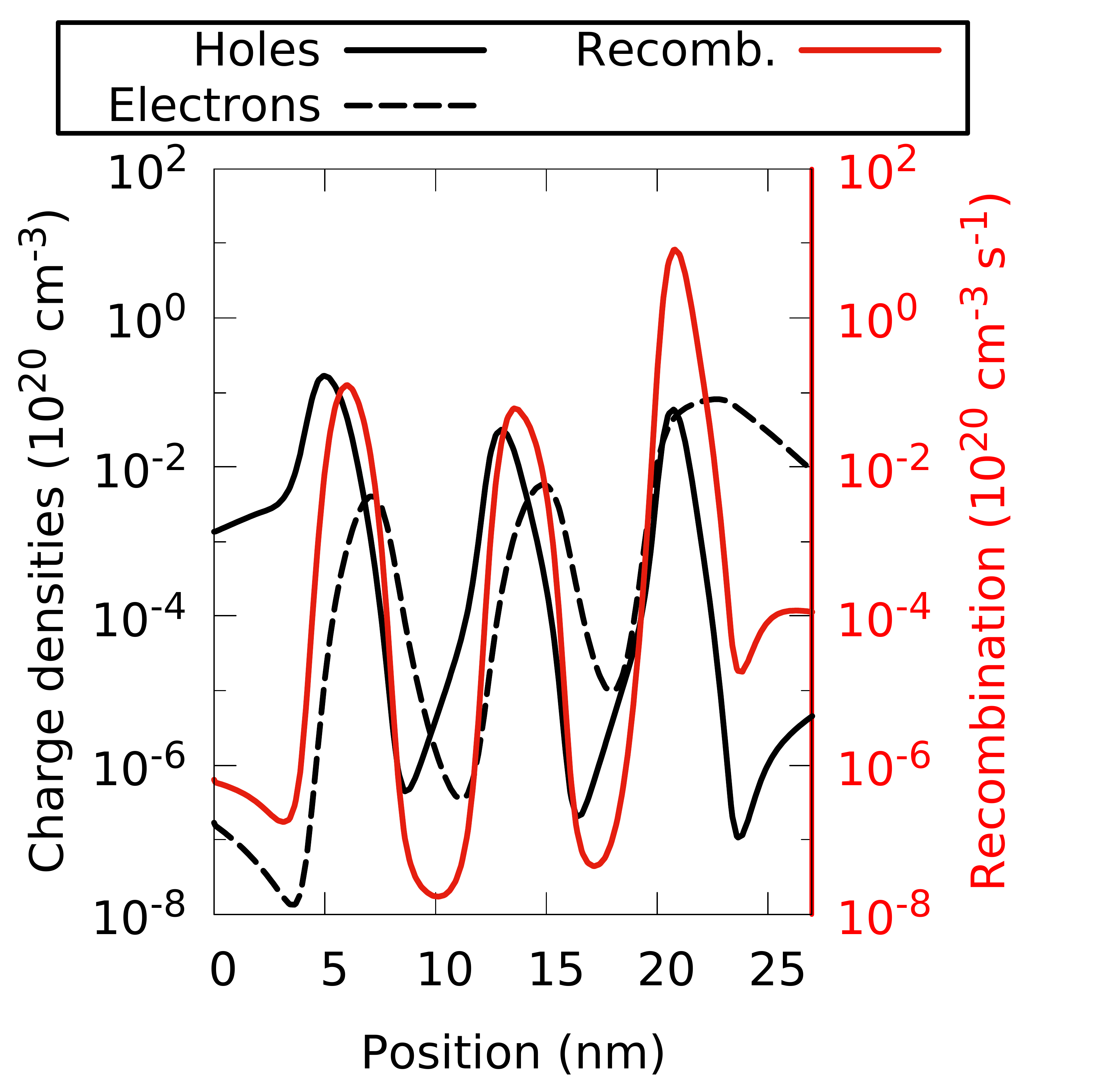} \\
        \hline
    \end{tabular}
    \caption{Hole density (black, solid), electron density (black, dashed), and radiative recombination rate (red, solid) averaged over each atomic plane along the transport direction.  Results from calculations building on (i) a virtual crystal approximation (top), (ii) a virtual crystal including quantum corrections via localization landscape theory (LLT) (center) and a (iii) random alloy description including LLT-based quantum corrections (bottom); the deep well is located at (a) the $n$-side (left), (b) the center (middle) and (c) the $p$-side (right). The data are shown on a log scale.}
    \label{fig:outofplaneDensities_log}
\end{figure*}

To examine the impact of random alloy fluctuations on the carrier distribution in an (In,Ga)N/GaN MQW stack, we start with a `standard' 1-D simulation approach that is widely applied in the literature. 
In a first step we begin with \NN{ } calculations and as outlined above, compare the results to our \texttt{ddfermi} data.
\subsubsection{\NN{} simulations}\label{subsubsec:NNresults}
To study how the presence of an EBL affects the ratio of radiative recombination $\varrho$, Eq.~(\ref{eq:RecombRatio}), the systems outlined in Section~\ref{sec:TheoryStructs} are simulated with and without a 20 nm Al$_{0.15}$Ga$_{0.85}$N EBL using \NN{}. The EBL is separated from the $p$-side QW (position 3 ($p$-side) in Fig.~\ref{fig:schematic_of_QWs}) by a 10 nm GaN barrier. Similar settings for an (Al,Ga)N EBL have been used in previous studies~\cite{LiPi2017}.
The \NN{ } calculated ratio of radiative recombination $\varrho$, 
when varying the position of the deep QW in the MQW stack, are depicted in Fig.~\ref{fig:recombratio}~(a). Turning first to the data without quantum corrections, we find that in the case of the employed 1-D VCA-like continuum-based description,  \emph{$\varrho$ is small when the deep QW is at the $n$-side (position 1 ($n$-side) Fig.~\ref{fig:schematic_of_QWs}) and larger when the deep well is at the p-side (position 3 ($p$-side) Fig.~\ref{fig:schematic_of_QWs})}. Thus, the 1-D model \emph{predicts the opposite trend} when compared to experiment~\cite{BGaPhysStatSol_2011}.
This trend is only slightly changed when including quantum corrections via a self-consistent Schr\"odinger-Poisson-DD model. In this case a much weaker dependence of the results on the position of the deep QW in the MQW stack is observed. However, even when including quantum corrections, the \NN{} results for $\varrho$ are not reflecting the experimentally observed behavior (see discussion above). Figure~\ref{fig:recombratio}~(a) reveals also that qualitatively the results do not depend on the presence of the EBL, indicating that for the structures considered, this feature of an LED is of secondary importance for the aims of this work. 

\subsubsection{\texttt{ddfermi} simulations}

Since we are also able to use the atomistic framework in a VCA setting, we compare our \texttt{ddfermi} results, cf. Fig.~\ref{fig:recombratio}~(b) (purple), with those from \NN{,} cf. Fig.~\ref{fig:recombratio}~(a) (purple, solid). We focus on structures which neglect the EBL as we have found above that it does not impact results in a VCA. In both \NN{ }and \texttt{ddfermi} a similar trend is found: the deep QW dominates recombination only when it is located at the $n$-side. This is illustrated further in Fig.~\ref{fig:percentagecontribution}~(a), which displays the contribution (in percent) to the radiative recombination rate from each QW (colors) in the MQW stack. The data are shown as a function of position of the deep QW in the MQW system. 
This confirms that it is always the QW which is closest to the $n$-doped side (position 1) that dominates the recombination process;
the $n$-side QW contributes $\approx 95\%$ when the deep QW is at position 1, $\approx 70\%$ when the deep QW is at position 2 or 3. Again, we stress that this is the \emph{opposite} trend to the experimental findings in Ref.~\cite{BGaPhysStatSol_2011}.

To shed more light on this result, the upper row in Figure~\ref{fig:outofplaneDensities_log} depicts the average hole (black, solid), electron (black, dashed) and radiative recombination (red) rate along the \textit{c}-axis when the deep QW (In$_{0.125}$Ga$_{0.875}$N well) is (a) closest to the $n$-side (position 1), (b) in the centre of the MQW stack (position 2) and (c) closest to the $p$-side (position 3). Focusing on the VCA data, Figs.~\ref{fig:outofplaneDensities_log}~(i)~(a-c), we see the cause of the dominant recombination from the $n$-side QW: the hole density is always high in this region, independent of which well is closest to the $n$-side. In particular, the $p$-side QW fails to capture holes effectively and consistently has the lowest hole density.
We note that a similar behavior is also found in the \NN{} calculations discussed in Sec.~\ref{subsubsec:NNresults}.



Given that our VCA \texttt{ddfermi} approach and \NN{ } treat (In,Ga)N as a homogeneous alloy that can be described by averaged material parameters which do not vary throughout the wells (no alloy fluctuations included), it allows us also to compare the implemented methods for quantum corrections in DD simulations. Here, as discussed above, \NN{} builds on the widely used Schr\"odinger-Poisson-DD model while \texttt{ddfermi} utilizes the recently developed LLT method. It has been discussed and shown in the literature that the LLT method can produce results in good agreement with the solution of the Schr\"odinger equation in the case of a 1-D effective mass approximation~\cite{FiPi2017,ChKe2020}.
Looking at Fig.~\ref{fig:recombratio}~(a) (green, solid) and Fig.~\ref{fig:recombratio}~(b) (green) we see that the results from our in-house developed \texttt{ddfermi}-based 3-D model, which employs LLT (3-D, \texttt{ddfermi}), are very similar to the standard self-consistent 1-D Schr\"odinger-Poisson-DD calculation underlying \NN{}. This gives confidence that our LLT treatment is providing a comparable description of the quantum corrections in the system.

Overall, Fig.~\ref{fig:recombratio}~(a) reveals that when including quantum corrections in the VCA calculations, the position of the deep QW has little impact on the ratio of the relative radiative recombination, $\varrho$.
From Fig.~\ref{fig:percentagecontribution}~(b) one can also gain more insight into this behavior and how quantum corrections impact the carrier distribution in the MQW stack. In the \emph{absence} of quantum corrections but utilizing a VCA, Fig.~\ref{fig:percentagecontribution}~(a), the well closest to the $n$-side dominates the relative radiative recombination ratio $\varrho$ independent of the position of the deep well in MQW systems. When \emph{including} quantum corrections this situation is now changed: the deep QW is now the dominant emitter independent of its position in the MQW stack.

This behavior becomes clear when analyzing the electron and hole densities as a function of the position of the deep well in the (In,Ga)N/GaN MQWs, as shown in Fig.~\ref{fig:outofplaneDensities_log}~(ii). Looking at the electron densities first, we find that electrons preferentially occupy the well closest to the $p$-side. This effect is enhanced when the deep QW is closest to the $p$-side (cf. Fig.~\ref{fig:outofplaneDensities_log}~(ii)~(c)). 
In our previous study on uni-polar electron transport~\cite{MiOD2021_JAP}, we have already seen that including quantum corrections leads to a softening of the potential barrier at the QW barrier interfaces. This in turn can lead to an increased electron current at a fixed bias point, when compared to a VCA system without LLT treatment, and thus the electrons can more easily `overshoot' the wells in the MQW system. As a consequence, a lower electron density in the well closest to the $n$-side is observed. Turning to the hole density, the situation is different. Here, we find that  holes preferentially populate the well closest to the $n$-side. Only when the deep QW is closest to the $p$-side, the hole density in this well is noticeably increased. 
However, when comparing the distribution of holes in the MQW as a function of the position of deep well in absence (Fig.~\ref{fig:outofplaneDensities_log}~(i)) and presence (Fig.~\ref{fig:outofplaneDensities_log}~(ii)) of quantum corrections, the results are not very different. This indicates that quantum corrections, at least when employing a VCA, are of secondary importance for the hole distribution. This finding is consistent with our previous results on uni-polar hole transport~\cite{OdFaOQE2022}, where we have discussed that due to the high effective hole mass and the small valence band offset, quantum corrections have a smaller impact on the hole transport when compared to electrons. As a consequence, the distribution of holes follows a similar pattern to that of the VCA where quantum corrections are neglected.
Finally, when looking at the ratio of radiative recombination $\varrho$, it is important to note that this quantity is not only determined by having both large electron and hole densities in the same well but also by their spatial overlap. As one can infer from Fig.~\ref{fig:outofplaneDensities_log}~(ii)~(a-c), the largest radiative recombination rate is always observed in the deepest well. This indicates also that the spatial overlap of electron and hole densities largest in the deep QW regardless of its position across the MQW system.
We stress again that even when including quantum corrections in the VCA calculations, the resulting trend in $\varrho$ is not reflecting the trend observed in experimental studies~\cite{BGaPhysStatSol_2011}.

\subsection{Impact of a random alloy fluctuations on the carrier transport in (In,Ga)N/GaN MQWs} \label{sec:ResultsFluctuations}

In the last step, we move away from the VCA description of the system and include, in addition to quantum corrections, also random alloy fluctuations in the calculations. Figure~\ref{fig:recombratio}~(b) (blue) shows that, and this time in line with the experimental results by Galler \emph{et. al}~\cite{BGaPhysStatSol_2011}, the deep QW only contributes significantly to the radiative recombination when it is \emph{closest to the $p$-side} (position 3). In fact, when including random alloy fluctuations in the calculations, the well closest to the $p$-side always has the largest contribution to total radiative recombination, as can be seen in Fig.~\ref{fig:percentagecontribution}~(c). 

To understand this behavior, Fig.~\ref{fig:outofplaneDensities_log}~(iii) depicts the electron and hole densities in the different wells as a function of the position of the deep well in the MQW systems. Looking at the electron density first, in comparison to the VCA calculations both including and excluding quantum corrections, random alloy fluctuations lead a reduction in electron density at the $n$-side QW. 
As discussed above and previously, quantum corrections can lead to increased electron transport, and including alloy fluctuations adds further to this effect due to the softening of the barrier at the well interfaces~\cite{MiOD2021_JAP}. 
As a consequence, the electrons can more easily `overshoot' the wells in the MQWs, which can also be seen in the increased electron density beyond the $p$-side QW when alloy fluctuations and quantum corrections are included. 
However, in comparison to the VCA result including quantum corrections, the electron density in the $p$-side well is only slightly affected by alloy fluctuations.

In contrast, hole densities in the $p$-side QW are more dramatically changed by alloy fluctuations. As Figs.~\ref{fig:outofplaneDensities_log}~(ii) and~\ref{fig:outofplaneDensities_log}~(iii) show, in comparison to the VCA description, alloy fluctuations lead to an increase in the hole density in the \emph{$p$-side QW} (position 3) even when the deep QW is closest to the $n$-side (position 1) or in the centre (position 2) of the MQW system. While the smoothing of the well barrier interface can increase hole transport, as in the case of electrons, there are now also alloy disorder induced localization effects to contend with. As discussed in our previous work, these localization effects are \emph{detrimental} to hole transport~\cite{OdFaOQE2022} and result in an increased hole density in the $p$-side QW. As a consequence, the well closest to the $p$-side dominates radiative recombination


We note that there is still a reasonable hole density present in the $n$-side QW (Fig.~\ref{fig:outofplaneDensities_log}~(iii)~(a-c)). In general, the distribution of carriers will also depend on the GaN barrier width and a 5 nm barrier is narrow enough to allow for some hole transport across the MQW~\cite{LiRyAPL2008}; a similar dependence of hole transmission on the barrier width has been seen in previous non-equilibrium Green's function studies~\cite{ODoLu2021}. Thus we expect that increasing the barrier width will mainly lead to a reduction of the hole density in the well furthest away from the $p$-side, but should to a lesser extent affect the hole density in the well closest to the $p$-side. Therefore, even for a larger barrier width than the here considered 5 nm, we expect that the recombination will still be dominated by the $p$-side QW.

We note that based on the VCA results we did not consider the EBL in the atomistic calculations. In general the EBL needs to be treated with an atomistic resolution. Previous studies of (Al,Ga)N barriers in uni-polar device settings have found that the impact of these barriers is lower than what is expected from a 1-D simulation for both electrons~\cite{BrMiSST2017} and holes~\cite{QwMoAPL2020}. Thus given that our VCA calculations show that the presence of the EBL is of secondary importance for our study, we expect a similar conclusion when treating the EBL with alloy fluctuations. Therefore, it is unlikely that the EBL impacts the here presented result, however this question may be targeted in future studies.

\section{Conclusions}\label{sec:conclusion}

In this work we apply a 3-D quantum corrected multiscale simulation framework to gain insight into the impact of random alloy fluctuations on the electron and hole distribution across the active region of an (In,Ga)N/GaN LED. 
To study the spatial distribution of carriers we have followed literature experimental studies~\cite{BGaPhysStatSol_2011} and analyzed the radiative recombination ratio in a multi-quantum well system, where one of the wells in the system has a higher indium content (deeper well) and its position is varied within the stack.

The here considered MQW systems are not only of interest for a comparison with experiment, they provide also the ideal opportunity to benchmark and validate results from our in-house developed 3-D multiscale simulation framework against commercially available software packages. To do so we treat the QWs in a virtual crystal approximation (VCA), to mimic the 1-D simulation widely used in the literature for (In,Ga)N QWs and implemented in the commercial software package \NN{}. In addition, this study allows us also to compare the different schemes to account for quantum corrections (localization landscape theory vs. Schr\"odinger-Poisson-DD simulations) in the simulations. Overall, this analysis showed very good agreement between results obtained from our in-house software and \NN{} (without random alloy fluctuation).

Equipped with this benchmarked model, our analysis reveals that including (random) alloy fluctuations in the calculations is vital for reproducing trends seen in experiment. More specifically, when using the widely employed virtual crystal approximation (VCA), the hole density in the well closest to the $p$-doped region of the device is significantly reduced compared to our atomistic random alloy calculation. As a consequence, and in contrast to the experiment, in VCA the well closest to the $p$-side contributes very little to the radiative recombination process, an effect that can be reduced by accounting for quantum corrections. 
While this leads to enhanced radiative recombination from the well closest to the $p$-side, at least when this well is the deep well, it still does not reflect the trends observed in the experimental studies. However, when including random alloy fluctuations and quantum corrections in our 3-D simulation framework, these effects lead to an increase in the hole density in the well closest to the $p$-side. Consequently, this well dominates the radiative recombination process in line with the experimental data. We note that in addition to quantum corrections and alloy fluctuations no further ingredients are required (e.g. multi-population model) to explain the experimentally observed trends.     
Therefore, our calculations highlight that alloy fluctuations are a key ingredient in simulations guiding the design of III-N based devices. Thus, the here developed model presents an ideal starting point for future calculations.

\begin{acknowledgements}
This work received funding from the Sustainable Energy Authority of Ireland and the Science Foundation Ireland (Nos. 17/CDA/4789, 12/RC/2276 P2 and 21/FFP-A/9014) and the Deutsche Forschungsgemeinschaft (DFG) under Germany’s Excellence Strategy EXC2046: MATH+, project AA2-15,  the Leibniz competition 2020 as well as the Labex CEMPI (ANR-11-LABX-0007-01).
\end{acknowledgements}


\bibliography{lit.bib}

\end{document}